\documentclass[letterpaper,english,american,reprint, onecolumn]{revtex4-1}
\usepackage[latin9]{inputenc}
\usepackage{color}
\usepackage{babel}
\usepackage{amsmath}
\usepackage{amsthm}
\usepackage{amssymb}
\usepackage{graphicx}
\usepackage[unicode=true,
 bookmarks=true,bookmarksnumbered=true,bookmarksopen=true,bookmarksopenlevel=1,
 breaklinks=false,pdfborder={0 0 1},backref=section,colorlinks=true]
 {hyperref}
\hypersetup{pdftitle={Linear stability analysis of detonations via numerical computation and dynamic mode decomposition},
 pdfauthor={Dmitry I. Kabanov and Aslan R. Kasimov},
 pdfsubject={detonation, linear stability, dynamic mode decomposition, numerics},
 pdfkeywords={detonation, DMD, stability}}

\makeatletter

\pdfpageheight\paperheight
\pdfpagewidth\paperwidth

\newcommand{\lyxdot}{.}

\@ifundefined{date}{}{\date{}}
\usepackage{newtxtext, newtxmath}

\usepackage{fancyhdr}
\usepackage{bm}
\usepackage{url}
\usepackage{color}

\usepackage{comment}

\usepackage{multirow}

\usepackage{dcolumn}

\usepackage[all]{hypcap}

\makeatother

\begin{document}

\title{Linear stability analysis of detonations\\
via numerical computation and dynamic mode decomposition}

\author{Dmitry I. Kabanov }

\affiliation{King Abdullah University of Science and Technology~\\
 Thuwal, Saudi Arabia}

\author{Aslan R. Kasimov}
\email{kasimov@lpi.ru}

\affiliation{P.\,N. Lebedev Physical Institute~\\
Russian Academy of Sciences~\\
Moscow, Russia}
\begin{abstract}
We introduce a new method to investigate linear stability of gaseous
detonations that is based on an accurate shock-fitting numerical integration
of the linearized reactive Euler equations with a subsequent analysis
of the computed solution via the dynamic mode decomposition. The method
is applied to the detonation models based on both the standard one-step
Arrhenius kinetics and two-step exothermic-endothermic reaction kinetics.
Stability spectra for all cases are computed and analyzed. The new
approach is shown to be a viable alternative to the traditional normal-mode
analysis used in detonation theory. 
\end{abstract}
\maketitle
\selectlanguage{english}%
\global\long\def\Eact{E_{\text{{act}}}}

\global\long\def\Nhrz{N_{1/2}}

\global\long\def\tolLambda{\text{tol}_{\lambda}}

\global\long\def\TFinal{T_{\text{f}}}

\global\long\def\rhoA{\rho_{\mathrm{a}}}

\global\long\def\rhoZnd{\bar{\rho}}

\global\long\def\uZnd{\bar{u}}

\global\long\def\UZnd{\bar{U}}

\global\long\def\pZnd{\bar{p}}

\global\long\def\TZnd{\bar{T}}

\global\long\def\lambdaZnd{\bar{\lambda}}

\global\long\def\rhoSZnd{\bar{\rho}_{\text{s}}}

\global\long\def\uSZnd{\bar{u}_{\text{s}}}

\global\long\def\USZnd{\bar{U}_{\text{s}}}

\global\long\def\pSZnd{\bar{p}_{\text{s}}}

\global\long\def\lambdaSZnd{\bar{\lambda}_{\text{s}}}

\global\long\def\rhoS{\rho_{\text{s}}}

\global\long\def\uS{u_{\text{s}}}

\global\long\def\pS{p_{\text{s}}}

\global\long\def\lambdaS{\lambda_{\text{s}}}

\global\long\def\dd{\,\text{d}}

\global\long\def\T{\text{T}}

\global\long\def\R{\mathbb{R}}

\global\long\def\norm#1{\|#1\|}

\global\long\def\rank{\text{rank}}

\global\long\def\ii{\mathrm{i}}

\global\long\def\diag{\text{diag}}

\global\long\def\abs#1{\left|#1\right|}

\global\long\def\od#1#2{\frac{\dd#1}{\dd#2}}

\global\long\def\pd#1#2{\frac{\partial#1}{\partial#2}}

\global\long\def\DD#1#2{\frac{\mathrm{D}#1}{\mathrm{D}#2}}

\global\long\def\DeltaX{\Delta x}

\global\long\def\DeltaT{\Delta t}
\selectlanguage{american}%

\section{Introduction}

Shock waves arise in great many areas in physics: compressible gas
dynamics, shallow-water flows, astrophysics, cosmology, quantum fluids,
and many others. They are often accompanied by time-dependent dynamic
phenomena related to instability of associated steady-state solutions.
Such instability may be caused by interactions between the flow, geometry
(boundaries, including the shock wave itself), and physical and chemical
processes taking place in the flowing matter (such as magnetic field
effects, chemical reactions, gravitation, quantum effects, and others)
\cite{dyakov1954,Erpenbeck64,bates2000d,fowles1973stability,swan1975shock,el2016dispersive,rezzolla1996stability,erpenbeck1962stability,kim2014nature,kontorovich1958concerning,kuznetsov1989stability,syrovatskii1959stability,foglizzo2007instability,fernandez2009stability}. 

Theoretical analysis of linear stability is usually based on the normal-mode
method whereby the underlying governing system of partial differential
equations is linearized, and the solution is represented as a certain
mode, the form of which depends on the geometry of the problem. The
mode is typically exponential in time as well as in those spatial
directions in which the steady state is uniform \cite{drazin2004hydrodynamic}.
In many interesting problems however, at least part of the steady-state
structure is not uniform spatially, and as a result the modal problem
becomes significantly more complicated. One typically must resort
then to a numerical solution of the associated system of ordinary
differential equations (ODEs) for the perturbation amplitudes. Finding
an accurate and robust algorithm for such numerical solution can be
challenging \cite{drazin2004hydrodynamic,schmid2012stability}. For
example, in a detonation problem considered in the present work, the
challenge stems from the complexity of reaction rate equations that
combined with the fluid equations, produce a system of ODEs that can
be stiff and/or singular. As a result, performing accurate calculations
of spectral properties becomes quite difficult \cite{short1998cellular,LeeStewart90,KasimovStewart02,sharpe1997linear,tumin2007initial,humpherys2010efficient,zumbrun2011stability,zumbrun2017recent,barker2017computing}.

Our goal in this paper is to introduce a new method for the calculation
of linear stability properties of shock waves and detonations. The
method consists of a direct numerical solution of a linearized system
of governing equations with a subsequent modal analysis of the computed
results for the presence of stable and unstable modes. One of the
novelties of our technique is that growth rates and frequencies of
the modes are extracted from the numerically computed time series
with the implementation of the dynamic mode decomposition (DMD) \cite{schmid2010dynamic,jovanovic2014sparsity,dawson2016characterizing}. 

The DMD is a powerful method to extract spectral information from
complex data coming from numerical or experimental studies of various
systems. It was introduced into analysis of fluid flows in \cite{schmid2008decomposition,schmid2010dynamic}.
Importantly, even though the DMD has originated and found its most
extensive use in fluid mechanics, the method is essentially model
independent as it deals directly with data, irrespective of their
origin. As such, it is of much interest in problems that involve large
amounts of data, and for which there is a need to extract some global
characteristic features. One can find the basic introduction to DMD
as well as the discussion of many of its modern applications in \cite{kutz2016dynamic}.
The latter begin with fluid mechanics, but extend to numerous other
fields involving ``big data'': atmospheric science, image processing,
epidemiology, neuroscience, financial markets and others. The problem
of understanding data is especially acute when there is no known model
that generates the data, for example, in the financial market or in
neuroscience. In contrast, in fluid mechanics the data are governed
by the Navier-Stokes or Euler equations, which are relatively well
established. 

Some of the recent developments of the method and its various applications
can be found in \cite{schmid2011application,schmid2011applications,chen2012variants,mezic2013analysis,tu2014dynamic,jovanovic2014sparsity,williams2015data,dawson2016characterizing,alekseev2016linear,brunton2016extracting,taira2017modal,rowley2017model,zhang2017evaluating,arbabi2017study,peitz2017koopman}.
There are interesting and deep mathematical roots of the method in
the dynamical systems theory in connection with the so-called Koopman
operator \cite{koopman1931hamiltonian,rowley2009spectral,mezic2013analysis},
which is an infinite-dimensional linear operator that represents a
given finite-dimensional but nonlinear operator. Koopman theory, which
is at the heart of DMD, can also be applied to analysis of various
partial differential equations (PDEs) and ODEs \cite{kutz2016koopman,alla2017nonlinear}.
A modification of DMD was recently used to characterize chaotic systems
\cite{brunton2017chaos}. In \cite{wynn2013optimal}, the authors
generalized DMD to optimal mode decomposition and showed that for
noise-free linear systems, DMD correctly identifies system eigenvalues
and hence is analogous to normal-mode analysis. Tu with co-authors
in \cite{tu2014dynamic} gave a relatively simple description of the
algorithm and showed that input data for DMD do not have to be sequential
time series with a uniform time step between data snapshots. They
also pointed out that snapshots must be sufficiently ``rich'' for
the DMD to extract dynamic modes correctly. In \cite{jovanovic2014sparsity},
a sparsity-promoting version of DMD algorithm is proposed to achieve
a balance between the quality of the extracted dynamics and the number
of modes required to represent the dynamics. Several researchers considered
sensitivity of DMD to the noise in input data. Duke with co-authors
\cite{duke2012error} considered synthetic wave shapes and carried
out a detailed study of errors in growth rates and frequencies under
varying parameters. The paper \cite{bagheri2014effects} demonstrated
that output DMD eigenvalues are damped in the presence of noise, while
\cite{dawson2016characterizing} proposed four modifications for the
DMD algorithm that alleviate noise effect. In \cite{kou2017improved,zhang2017evaluating},
criteria for choosing the rank of a low-order approximation are proposed,
with \cite{kou2017improved} suggesting to determine relevant modes
through its time-dependent amplitude, while \cite{zhang2017evaluating}
proposing to choose mode eigenfunctions on the basis of how good they
approximate the corresponding Koopman operator eigenfunctions.

One can find applications of DMD in combustion problems as well. For
example, the authors of \cite{massa2012dynamic} used DMD to investigate
the dynamics of multidimensional detonations by analyzing the flow
structures computed from the reactive compressible Navier-Stokes equations.
We note that the analysis \cite{massa2012dynamic} was concerned with
full nonlinear dynamics. In contrast, in this paper we investigate
linear stability of one-dimensional reactive Euler equations. Thus
our system is less general, however we provide more detailed and extensive
analysis of the dynamics of the chosen model. In \cite{richecoeur2012dmd},
DMD is applied to experimental study of combustion instability by
stacking different flow observables (pressure, velocity, etc.).

In this work, we solve the linearized reactive Euler equations using
a high-order space and time integration algorithm that is based on
the shock fitting to avoid numerical shock-capturing errors\textbf{
}\cite{kasimov2004dynamics,HenrickAslamPowers2006,Taylor-Kasimov-Stewart-CTM09,hairer1993solving}.
Using the new method, we calculate stability spectra and neutral boundaries
for the problem and compare them with prior results found by the method
of normal modes \cite{LeeStewart90,short1998cellular,sharpe1997linear}.
The computations are carried out for the standard one-step Arrhenius
heat release and also for a two-step heat release with one exothermic
and one endothermic reactions (the so-called ``pathological detonation''
\cite{sharpe2000one,FickettDavis2011}). 

For the standard one-step chemistry, we find complete agreement with
the previously published results. For the two-step chemistry of the
type $\text{A}\to\text{B}\to\text{C}$ where the first reaction is
exothermic and the second is endothermic, we analyze both types (subsonic-supersonic
and subsonic-subsonic) of the steady-state solutions that exist. For
the subsonic-supersonic steady state, we qualitatively compare our
result with the results of Sharpe \cite{sharpe1999linearpathological}
and again good agreement is found. Additionally, we calculate the
stability of the subsonic-subsonic steady state and show that the
eigenfunctions of the two steady-state solutions are the same between
the shock and the sonic point, even though they are different downstream.
Perhaps more importantly, our method of solution of the linear stability
problem is essentially at the same level of complexity as that for
the single-step chemistry case. In contrast, the method in \cite{sharpe1999linearpathological}
requires an intricate asymptotic analysis of perturbations in the
neighborhood of the sonic locus such that an unbounded component of
the solution can be identified and discarded. Imposing such a boundedness
condition is generally the least understood and most difficult part
of normal-mode stability analysis of detonations. The methodology
proposed in the present work aims at circumventing this difficulty
albeit at the expense of more extensive numerical calculations.

The remainder of the paper is organized as follows. In Section \ref{sec:Governing-equations},
we introduce the governing system of reactive Euler equations. In
Section \ref{sec:algorithm}, we present the algorithms for the numerical
integration of the linearized governing equations and for the processing
of the generated time series of the shock velocity. In Section \ref{sec:results},
calculations of instability of detonations with various rate functions
are presented. In Section \ref{sec:Conclusions}, we discuss the results
and possible extensions of the ideas presented in this work.

\section{Reactive Euler equations \label{sec:Governing-equations}}

Our analysis of detonation stability is based on the one-dimensional
reactive Euler equations for a perfect gas that has a constant ratio
of specific heats and reacts following either a single-step Arrhenius
law or a two-step mechanism of the type $\text{A}\to\text{B}\to\text{C}$.
The steady traveling-wave solutions for these two cases are investigated
for linear stability by means of direct numerical integration of the
linearized Euler equations with a subsequent analysis of the resultant
time series of the shock speed using the DMD algorithm. In this section,
we introduce the governing equations, their steady-state solutions
and linearization about the steady state. 

\subsection{One-step Arrhenius kinetics }

First, we present the governing equations for the case of one-step
irreversible reaction 
\begin{equation}
\text{A}\to\text{B}.
\end{equation}
The progress of this reaction is followed by a variable $\lambda$
such that the mass fraction of $\text{A}$ is $1-\lambda$ and that
of $\text{B}$ is $\lambda$. 

The wave is assumed to propagate from left to right with speed $D$,
and in the reference frame attached to the lead shock, the governing
equations are written as \cite{kasimov2004dynamics}:
\begin{align}
\rho_{t}+\left(\rho\left(u-D\right)\right)_{x} & =0,\label{eq:mass}\\
\left(\rho u\right)_{t}+\left(\rho u\left(u-D\right)+p\right)_{x} & =0,\label{eq:momentum}\\
\left(\rho e\right)_{t}+\left(\rho\left(u-D\right)e+pu\right)_{x} & =0,\label{eq:energy}\\
\left(\rho\lambda\right)_{t}+\left(\rho\left(u-D\right)\lambda\right)_{x} & =\rho\omega.\label{eq:reaction}
\end{align}
where $\rho$, $u$, $p$, $e$, $\omega$ are density, flow velocity,
pressure, total specific energy, and reaction rate, respectively.
The specific energy is $e=e_{\text{i}}+u^{2}/2$ with constitutive
relations of a calorically perfect gas: 
\begin{equation}
e_{\text{i}}=\frac{pv}{\gamma-1}-Q\lambda,\qquad pv=RT,\label{eq:e_i}
\end{equation}
where $\gamma$ is the polytropic index, $v=1/\rho$ specific volume,
$Q$ chemical heat release, $R$ the universal gas constant, $T$
temperature. The reaction rate is taken here to follow simple-depletion
Arrhenius law:
\begin{equation}
\omega=k\left(1-\lambda\right)\exp\left(-\frac{E\rho}{p}\right),\label{eq:Arrhenius}
\end{equation}
where $k$ is rate constant and $E$ the activation energy.

The numerical solution of the governing equations (\ref{eq:mass}\textendash \ref{eq:reaction})
in the shock-attached frame requires an evolution equation for $D$
which is present explicitly in the equations. For this purpose, we
use the equation derived in \cite{Taylor-Kasimov-Stewart-CTM09} by
specializing it to one dimension: 
\begin{equation}
\frac{dM}{dt}=s,\label{eq:shock-change}
\end{equation}
where $M=-\rho_{\text{a}}D$ is the normal mass flux into the shock,
$\rho_{\text{a}}$ is the upstream density, and 
\begin{align}
s & =\frac{1}{A_{0}}\left(R_{\text{s}}-A_{\text{s}}\right), & A_{0} & =\frac{2}{\gamma+1}Mv_{\text{a}}\left(3+\frac{\gamma p_{\text{a}}\rho_{\text{a}}}{M^{2}}\right),\label{eq:s-A0}\\
R_{\text{s}} & =Q(\gamma-1)\rho_{\text{s}}\omega_{\text{s}}, & A_{\text{s}} & =\rho_{\text{s}}(c_{\text{s}}^{2}-U_{\text{s}}^{2})u_{x}|_{\text{s}},\label{eq:Rs-As}
\end{align}
where $c=\sqrt{\gamma p/\rho}$ is the sound speed, $U=u-D$ is the
flow velocity in the shock-attached frame, subscript ``s'' denotes
the shock state and subscript ``a'' the ambient (upstream) state.
Note that in (\ref{eq:s-A0}-\ref{eq:Rs-As}) everything depends on
the shock state through the Rankine-Hugoniot conditions except for
the velocity gradient $u_{x}|_{\text{s}}$ at the shock, which must
be determined numerically by solving for the flow downstream. 

The Rankine\textendash Hugoniot conditions at $x=0$ give all the
state variables in terms of the mass flux (or, equivalently, the shock
speed): 
\begin{align}
p_{\text{s}} & =-\frac{\gamma-1}{\gamma+1}p_{\text{a}}+\frac{2v_{\text{a}}}{\gamma+1}M^{2}, & v_{\text{s}} & =\frac{\gamma-1}{\gamma+1}v_{\text{a}}+\frac{2}{\gamma+1}\frac{\gamma p_{\text{a}}}{M^{2}}, & U_{\text{s}} & =Mv_{\text{s}},\label{RH-conditions}
\end{align}
and $\lambda_{\text{s}}=0$, $\rho_{\text{s}}=1/v_{\text{s}}$.

Next, we introduce the vector of state variables $z={(\rho,u,p,\lambda)}^{\mathrm{T}}$
and linearize it as well as the shock speed as $z(x,t)=\bar{z}(x)+\epsilon z'(x,t)$,
$D(t)=\bar{D}+\epsilon\psi'(t)$, with $0<\epsilon\ll1$, the bar
denoting the steady-state, and the prime denoting the perturbation
amplitudes. Upon linearization of the governing equations (\ref{eq:mass}\textendash \ref{eq:reaction}),
we obtain 
\begin{equation}
z'_{t}+A(\bar{z})z'_{x}+B(\bar{z})z'-\frac{d\bar{z}}{dx}\psi'=0,\label{linearized-system}
\end{equation}
where 
\begin{equation}
A=\begin{bmatrix}\bar{U} & \bar{\rho} & 0 & 0\\
0 & \bar{U} & 1/\bar{\rho} & 0\\
0 & \gamma\bar{p} & \bar{U} & 0\\
0 & 0 & 0 & \bar{U}
\end{bmatrix},\quad B=\begin{bmatrix}\frac{d\bar{u}}{dx} & \frac{d\bar{\rho}}{dx} & 0 & 0\\[0.3em]
-\frac{1}{\bar{\rho}^{2}}\frac{d\bar{p}}{dx} & \frac{d\bar{u}}{dx} & 0 & 0\\[0.3em]
C(\bar{\rho}\bar{\omega}_{\rho}+\bar{\omega}) & \frac{d\bar{p}}{dx} & \gamma\frac{d\bar{u}}{dx}+C\bar{\rho}\bar{\omega}_{p} & C\bar{\rho}\bar{\omega}_{\lambda}\\[0.3em]
-\bar{\omega}_{\rho} & \frac{d\bar{\lambda}}{dx} & -\bar{\omega}_{p} & -\bar{\omega}_{\lambda}
\end{bmatrix},\label{eq:AB}
\end{equation}
$C=-(\gamma-1)Q$, and subscripts $\rho$, $p$ and $\lambda$ denote
partial derivatives.

The linearization of the Rankine\textendash Hugoniot conditions and
the shock-evolution equation yields: 
\begin{equation}
p'_{\text{s}}=\frac{4v_{\text{a}}\bar{M}}{\gamma+1}M',\quad v'_{\text{s}}=-\frac{4\gamma p_{\text{a}}}{\left(\gamma+1\right)\bar{M}^{3}}M',\quad U'_{\text{s}}=\bar{M}v_{\text{s}}'+\bar{v}_{\text{s}}M',\quad\rho_{\text{s}}'=-\frac{v'_{\text{s}}}{\bar{v}_{\text{s}}^{2}},\quad\lambda_{\text{s}}'=0,\label{RH-conditions-linearized}
\end{equation}
\begin{equation}
\frac{dM'}{dt}=s',\label{shock-evolution-equation}
\end{equation}
where 
\begin{equation}
s'=\frac{1}{\bar{A}_{0}}(R_{\text{s}}'-A_{\text{s}}')-\frac{A_{0}'}{\bar{A}_{0}^{2}}(\bar{R}_{\text{s}}-\bar{A}_{\text{s}}),
\end{equation}
\begin{equation}
A_{0}'=\frac{2}{\gamma+1}\frac{3v_{\text{a}}\bar{M}^{2}-\gamma p_{\text{a}}}{\bar{M}^{2}}M',\quad R_{\text{s}}'=Q(\gamma-1)\left.\left[\left(\bar{\rho}\bar{\omega}_{\rho}+\bar{\omega}_{\text{}}\right)\rho'+\bar{\rho}\bar{\omega}_{p}p_{\text{}}'\right]\right|_{\text{s}},
\end{equation}
\begin{equation}
A_{\text{s}}'=\left.\left[\rho\frac{d\bar{u}}{dx}\left(\gamma\left(\bar{p}v'+\bar{v}p'\right)-2\bar{U}U'\right)+\left(\bar{c}^{2}-\bar{U}^{2}\right)\frac{d\bar{u}}{dx}\rho'+\bar{\rho}\left(\bar{c}^{2}-\bar{U}^{2}\right)u'_{x}\right]\right|_{\text{s}}
\end{equation}
with all the spatially-dependent quantities on the right-hand sides
of $R'_{\text{s}}$ and $A'_{\text{s}}$ evaluated at the shock.

The base steady-state traveling wave is the ZND solution (after Zel'dovich
\cite{Zeldovich1940}, von Neumann \cite{vonNeumann1942} and D\"{o}ring
\cite{Doering1943}) given in terms of $\bar{\lambda}$ and $\bar{D}$
by \cite{Taylor-Kasimov-Stewart-CTM09}: 
\begin{equation}
\bar{v}=\frac{\gamma}{\gamma+1}\frac{\bar{P}}{\bar{M}^{2}}(1-\delta),\quad\bar{p}=\bar{P}-\bar{M}^{2}\bar{v},\quad\bar{U}=\bar{M}\bar{v},\label{eq:steady-state}
\end{equation}
where 
\begin{equation}
\delta=\sqrt{1-\frac{h\bar{M}^{2}}{\bar{P}^{2}}\left(\bar{H}+Q\bar{\lambda}\right)},\quad h=\frac{2(\gamma^{2}-1)}{\gamma^{2}},\label{eq:delta}
\end{equation}
and $\bar{P}=p_{\text{a}}+\rho_{\text{a}}\bar{D}^{2}$, $\bar{H}=e_{\text{a}}+p_{\text{a}}/\rho_{\text{a}}+\bar{D}^{2}/2$.
To obtain the solution in terms of $x$, the steady-state rate equation
\begin{equation}
\frac{d\bar{\lambda}}{dx}=\frac{\bar{\omega}}{\bar{U}}\label{eq:dlambda_dx}
\end{equation}
is integrated starting at the shock with the initial condition $\bar{\lambda}_{\text{s}}=0$
and auxiliary relations~\eqref{eq:steady-state} and \eqref{eq:delta}.

For self-sustained detonations with the Arrhenius kinetics~\eqref{eq:Arrhenius}
the flow becomes sonic relative to the shock at the end of the reaction
zone $\lambda=1$, which is located at $x\to-\infty$. Then $\delta=0$
at $\lambda=1$ \cite{Kasimov-phd,KasimovStewart05}, and it follows
that $\bar{D}=D_{\mathrm{CJ}}$ where
\begin{equation}
D_{\mathrm{CJ}}=\sqrt{c_{\text{a}}^{2}+q}+\sqrt{q},\qquad q=\frac{(\gamma^{2}-1)Q}{2},\qquad c_{\text{a}}^{2}=\frac{\gamma p_{\text{a}}}{\rho_{\text{a}}}.\label{eq:D_CJ}
\end{equation}

The pre-exponential factor $k$ defines the length scale and is chosen
following the standard practice such that the half-reaction length
is unity:
\begin{equation}
k=\int_{0}^{1/2}\frac{\bar{U}}{\left(1-\bar{\lambda}\right)\exp\left(-E\bar{\rho}/\bar{p}\right)}\,d\bar{\lambda}.\label{eq:k}
\end{equation}

\subsection{Two consecutive reactions}

Now consider a simple example of a multi-step chemistry that consists
of two consecutive reactions: 
\begin{equation}
\text{A}\overset{k_{1}}{\to}\text{B}\overset{k_{2}}{\to}\text{C},
\end{equation}
where the first reaction is exothermic and the second endothermic.
This is often referred to as the case of a ``pathological'' detonation
\cite{FickettDavis2011}. The unfortunate name comes from the fact
that the detonation in this case propagates faster than expected,
the expectation based on the (wrong) thinking that it is the overall
heat release that controls the detonation speed. In reality, it is
not the overall but rather the maximum heat release\textemdash which
is larger\textemdash that controls the wave speed. This should become
clear from the analysis of the steady-state structure that follows. 

Let the reactions have rate constants $k_{1}$ and $k_{2}$, heat
releases $Q_{1}>0$ and $Q_{2}<0$, and activation energies, $E_{1}$
and $E_{2}$, for the first ($\text{A}\to\text{B}$), and second,
($\text{B}\to\text{C}$), reactions, respectively. Then, if $X_{A}$,
$X_{B}$, and $X_{C}$ denote the mass fractions of the species, the
kinetic equations for the reactions are 
\begin{eqnarray}
\dot{X}_{A} & = & -k_{1}X_{A}e^{-E_{1}/RT},\\
\dot{X}_{B} & = & k_{1}X_{A}e^{-E_{1}/RT}-k_{2}X_{B}e^{-E_{2}/RT},\\
\dot{X}_{C} & = & k_{2}X_{B}e^{-E_{2}/RT}.
\end{eqnarray}
Adding these equations and noting that initially $X_{A}=1$, $X_{B}=X_{C}=0$,
we find that $X_{A}+X_{B}+X_{C}=1$. We introduce the reaction progress
variables for each reaction as $\lambda_{1}=1-X_{A}$ and $\lambda_{2}=X_{C}$
with $\lambda_{1}$ and $\lambda_{2}$ varying from 0 at the shock
to 1 at the end of the reaction. Then we obtain the following rate
equations 
\begin{align}
\left(\rho\lambda_{1}\right)_{t}+\left(\rho(u-D)\lambda_{1}\right)_{x} & =\rho\omega_{1},\\
\left(\rho\lambda_{2}\right)_{t}+\left(\rho(u-D)\lambda_{2}\right)_{x} & =\rho\omega_{2},
\end{align}
with $\omega_{1}=k_{1}\left(1-\lambda_{1}\right)e^{-E_{1}/RT}$ and
$\omega_{2}=k_{2}\left(\lambda_{1}-\lambda_{2}\right)e^{-E_{2}/RT}$.
These equations replace the rate equation \eqref{eq:reaction}. The
specific internal energy becomes $e_{i}=pv/\left(\gamma-1\right)-\left(Q_{1}\lambda_{1}+Q_{2}\lambda_{2}\right)$.
The steady-state formulas retain their form with $Q\lambda$ replaced
by $Q_{1}\lambda_{1}+Q_{2}\lambda_{2}$. In particular,
\begin{equation}
\delta=\sqrt{1-\frac{h\bar{M}^{2}}{\bar{P}^{2}}\left(H+Q_{1}\bar{\lambda}_{1}+Q_{2}\bar{\lambda}_{2}\right)}.\label{eq:delta-2step}
\end{equation}
Recall that $\delta=0$ when the flow becomes sonic relative to the
shock \cite{Kasimov-phd}. Then, for $\delta$ to have a derivative
that is everywhere bounded we require that 
\begin{equation}
Q_{1}\frac{d\bar{\lambda}_{1}}{dx}+Q_{2}\frac{d\bar{\lambda}_{2}}{dx}=0\text{ whenever }\delta=0.\label{eq:regularity}
\end{equation}
This means that the heat release has a maximum at the sonic point.
The solution of $\delta=0$ yields the same result for $\bar{D}$
as before, i.e., 
\begin{equation}
\bar{D}=\sqrt{c_{\text{a}}^{2}+\frac{\gamma^{2}-1}{2}\left(Q_{1}\bar{\lambda}_{1}^{*}+Q_{2}\bar{\lambda}_{2}^{*}\right)}+\sqrt{\frac{\gamma^{2}-1}{2}\left(Q_{1}\bar{\lambda}_{1}^{*}+Q_{2}\bar{\lambda}_{2}^{*}\right)}\label{eq:D-2step}
\end{equation}
with star denoting the sonic locus. The first of the conditions \eqref{eq:regularity}
(known as the thermicity condition \cite{FickettDavis2011}) can be
written using the steady-state rate equations 
\begin{eqnarray}
\bar{U}\frac{d\bar{\lambda}_{1}}{dx}=\bar{\omega}_{1}, & \quad & \bar{U}\frac{d\bar{\lambda}_{2}}{dx}=\bar{\omega}_{2}\label{eq:steady-rates-2step}
\end{eqnarray}
as 
\begin{equation}
Q_{1}\bar{\omega}_{1}^{*}+Q_{2}\bar{\omega}_{2}^{*}=0.\label{eq:thermicity-2step}
\end{equation}
It follows from this equation that 
\begin{equation}
\bar{\lambda}_{2}^{*}=\bar{\lambda}_{1}^{*}+\frac{k_{1}Q_{1}}{k_{2}Q_{2}}\left(1-\bar{\lambda}_{1}^{*}\right)\exp\left(-\frac{E_{1}-E_{2}}{R\bar{T}^{*}}\right),\label{eq:lambda2star}
\end{equation}
where $\bar{T}^{*}$ is some complicated function of $\bar{\lambda}_{1}^{*}$,
$\bar{\lambda}_{2}^{*}$ and $\bar{D}$ that can be found from the
steady-state relations. 

Generally, the sonic state and the detonation speed are calculated
as follows. For a chosen $\bar{D}$, one solves \eqref{eq:steady-rates-2step}
starting from the shock to the sonic point, $x_{*}$, where the sonic
conditions \eqref{eq:D-2step} and \eqref{eq:thermicity-2step} must
be satisfied. For arbitrary $\bar{D}$, these conditions will not
be satisfied, and instead the steady-state solution will blow up.
But for special $\bar{D}$ (one or more), the solution will pass continuously
(however, not necessarily smoothly) through $x_{*}$ to the equilibrium
state at $x\to-\infty$. This is a general numerical procedure that
can be followed to determine the detonation speed $\bar{D}$.

We however restrict our attention in this work to the case $E_{1}=E_{2}$
which allows for a more explicit analytical calculation that will
facilitate both the theory and numerical computations that follow.
Expression \eqref{eq:lambda2star} simplifies then to 
\begin{equation}
\bar{\lambda}_{2}^{*}=\bar{\lambda}_{1}^{*}+\frac{k_{1}Q_{1}}{k_{2}Q_{2}}\left(1-\bar{\lambda}_{1}^{*}\right).\label{eq:lambdastar-12}
\end{equation}
Upon division of the second equation in \eqref{eq:steady-rates-2step}
by the first, we obtain 
\begin{gather}
\frac{d\bar{\lambda}_{2}}{d\bar{\lambda}_{1}}=k_{r}\frac{\bar{\lambda}_{1}-\bar{\lambda}_{2}}{1-\bar{\lambda}_{1}},\label{eq:lam2lam1}
\end{gather}
where $k_{r}=k_{2}/k_{1}$. The linear equation \eqref{eq:lam2lam1}
can be integrated to yield 
\[
\bar{\lambda}_{2}=\begin{cases}
1-\frac{\left(1-\bar{\lambda}_{1}\right)^{k_{r}}-k_{r}\left(1-\bar{\lambda}_{1}\right)}{1-k_{r}}, & k_{r}\neq0,\\
1-\left(1-\bar{\lambda}_{1}\right)\left(1-\ln\left(1-\bar{\lambda}_{1}\right)\right), & k_{r}=1.
\end{cases}
\]

Thus, to find $\bar{\lambda}_{1}^{*}$, we need to substitute one
of these expressions for $\bar{\lambda}_{2}$ into \eqref{eq:lambdastar-12}
and then solve the resultant nonlinear equation for $\bar{\lambda}_{1}^{*}$,
which requires a numerical iterative procedure. If we further restrict
the case to $k_{r}=1$, we find explicitly that
\begin{equation}
\left(1-\bar{\lambda}_{1}^{*}\right)\left[\ln\left(1-\bar{\lambda}_{1}^{*}\right)-\frac{Q_{1}}{Q_{2}}\right]=0,
\end{equation}
from which it follows that either $\bar{\lambda}_{1}^{*}=\bar{\lambda}_{1}^{e}=1$,
which is the equilibrium state, or 
\begin{equation}
\bar{\lambda}_{1}^{*}=1-\exp\left(Q_{1}/Q_{2}\right),
\end{equation}
which is the embedded sonic state. 

For the equilibrium case, $\bar{\lambda}_{1}^{e}=\bar{\lambda}_{2}^{e}=1$,
the wave speed can be found from $\delta_{e}=0$, and it is the same
as given in \eqref{eq:D_CJ} with $Q=Q_{1}+Q_{2}$. This is sometimes
called an ``equilibrium CJ solution''. We emphasize however that
even though formally we can calculate this velocity, in reality the
governing equations have no solution at this speed, and therefore
it makes no sense to talk about ``equilibrium CJ solution''. The
reason for the lack of solution can be seen from \eqref{eq:delta-2step}.
The expression $Q_{1}\bar{\lambda}_{1}+Q_{2}\bar{\lambda}_{2}$ goes
from $0$ at the shock to $Q_{1}+Q_{2}$ at equilibrium, but along
the way, it has a maximum at the embedded sonic state. Therefore,
if $\delta=0$ at $\bar{\lambda}_{1}=\bar{\lambda}_{2}=1$, then the
expression under the square root in \eqref{eq:delta-2step} would
have to take on negative values in some part of the reaction zone,
which is of course impossible. Thus, we must exclude the ``equilibrium
CJ solution'' from consideration.

The second case gives an embedded sonic locus. The flow is divided
then by the sonic point into two parts. Between the shock and the
sonic point, the flow is always subsonic relative to the shock. Between
the sonic point and equilibrium point, the flow can be either subsonic
or supersonic: flow is supersonic if expression \eqref{eq:delta-2step}
is used as is, while subsonic if the sign of the right-hand side of
\eqref{eq:delta-2step} is changed to negative \cite{Kasimov-phd}.
Thus we obtain two steady-state solutions with the embedded sonic
locus. Stability of both will be investigated below. 

The linearized perturbation system is the same as before, but with
$z={(\rho,u,p,\lambda_{1},\lambda_{2})}^{\mathrm{T}}$ and with the
following new matrices of coefficients (with $C_{i}=-(\gamma-1)Q_{i}$
and $i=1,2$):
\begin{align}
A=\begin{bmatrix}\bar{U} & \bar{\rho} & 0 & 0 & 0\\
0 & \bar{U} & 1/\bar{\rho} & 0 & 0\\
0 & \gamma\bar{p} & \bar{U} & 0 & 0\\
0 & 0 & 0 & \bar{U} & 0\\
0 & 0 & 0 & 0 & \bar{U}
\end{bmatrix},\quad & B=\begin{bmatrix}\frac{d\bar{u}}{dx} & \frac{d\bar{\rho}}{dx} & 0 & 0 & 0\\[0.3em]
-\frac{1}{\bar{\rho}^{2}}\frac{d\bar{p}}{dx} & \frac{d\bar{u}}{dx} & 0 & 0 & 0\\[0.3em]
\sum_{i}C_{i}(\bar{\rho}\bar{\omega}_{i\rho}+\bar{\omega}_{i}) & \frac{d\bar{p}}{dx} & \gamma\frac{d\bar{u}}{dx}+\bar{\rho}\sum_{i}C_{i}\bar{\omega}_{ip} & \bar{\rho}\sum_{i}C_{i}\bar{\omega}_{i\lambda_{1}} & \bar{\rho}\sum_{i}C_{i}\bar{\omega}_{i\lambda_{2}}\\[0.3em]
-\bar{\omega}_{1\rho} & \frac{d\bar{\lambda}_{1}}{dx} & -\bar{\omega}_{1p} & -\bar{\omega}_{1\lambda_{1}} & -\bar{\omega}_{1\lambda_{2}}\\
-\bar{\omega}_{2\rho} & \frac{d\bar{\lambda}_{2}}{dx} & -\bar{\omega}_{2p} & -\bar{\omega}_{2\lambda_{1}} & -\bar{\omega}_{2\lambda_{2}}
\end{bmatrix}.\label{eq:AB-2reaction}
\end{align}

For the shock evolution equation we obtain 
\begin{eqnarray}
\bar{R}_{\text{s}} & = & (\gamma-1)\rho_{\text{s}}\sum_{i}Q_{i}\bar{\omega}_{i\text{s}},\\
R_{\text{s}}' & = & (\gamma-1)\sum_{i}Q_{i}\left.\left[\left(\bar{\rho}\bar{\omega}_{i\rho}+\bar{\omega}_{i}\right)\rho'+\bar{\rho}\bar{\omega}_{ip}p'\right]\right|_{\text{s}}.
\end{eqnarray}
All the other expressions remain as before.

The equations are rescaled using the so-called Erpenbeck scales such
that $\rho_{\text{a}}=p_{\text{a}}=1$, velocity is scaled by $\sqrt{p_{\text{a}}/\rho_{\text{a}}}$,
spatial scale is that of the half-reaction length. With these scales,
equations keep their form. Activation energy and heat release are
rescaled by the ambient temperature multiplied by $R$.

\selectlanguage{english}%

\section{The numerical solution procedure\label{sec:algorithm}}

To determine stability properties of self-sustained detonations for
any particular set of physical parameters $Q$, $E$, and $\gamma$,
we solve an initial boundary value problem for (\ref{linearized-system}\textendash \ref{shock-evolution-equation})
numerically to see if its solution grows or decays in time. In each
simulation, after an initialization stage consisting of grid partitioning
and the evaluation of ZND quantities on the grid, the linear system
is integrated in time with simultaneous recording of the time series
of the perturbation of detonation velocity, $\psi'$, and then the
time series are post-processed using DMD to extract a stability spectrum.
Each of these stages is described in detail below.

\begin{figure}
\begin{centering}
\includegraphics{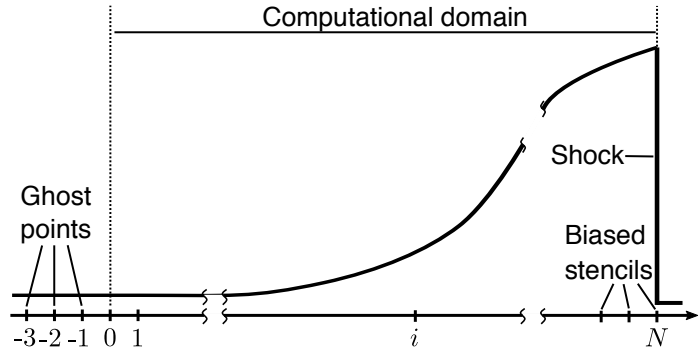}
\par\end{centering}
\caption{Schematics of the numerical grid.\label{fig:algorithm:grid}}
\end{figure}

With the rate function \eqref{eq:Arrhenius}, the reaction zone is
formally infinite. For numerical simulations, this length must be
truncated to some finite value $\mathcal{L}$. We determine $\mathcal{L}$
by the closeness of the reaction progress variable of the ZND solution
$\lambdaZnd$ to unity at the end of the reaction zone as measured
by some tolerance $\tolLambda$. Once $\tolLambda$ is specified,
we compute $\mathcal{L}$ using
\begin{equation}
\mathcal{L}=\lceil I\rceil\text{ with }I=\int_{0}^{1-\tolLambda}\frac{\UZnd}{k\left(1-\lambdaZnd\right)\exp\left(-\rhoZnd E/\pZnd\right)}\dd\lambdaZnd,
\end{equation}
where $k$ is computed from \eqref{eq:k}. The computational domain
is then discretized using a uniform grid $\left\{ x_{i}\right\} $,
$i=0,\dots,N$ with $N=\Nhrz\mathcal{L}$; $x_{0}=-\mathcal{L}$ and
$x_{N}=0$ being left and right boundary points, respectively. Here
$\Nhrz$ denotes the number of grid points per unit length (to remind,
the unit length is the distance from the shock to the point where
$\lambda=1/2$, hence the index $1/2$). The spatial step of the grid
is $\DeltaX=1/\Nhrz$. Schematically, the grid is shown in Figure
\ref{fig:algorithm:grid}.

The steady-state ZND profiles are computed on the grid using (\ref{eq:steady-state}-\ref{eq:dlambda_dx})
via the numerical integrator VODE \cite{brown1989vode} with the BDF
method of fifth order with absolute and relative tolerances set to
$10^{-15}$. Also, all other ZND quantities that enter matrices $A(\bar{z})$
and $B(\bar{z})$ in \eqref{eq:AB} are computed.

\subsection{Integration of the linearized system\label{subsec:linearized-system}}

After the initialization stage, an unsteady solution of the linearized
system is computed by the method of lines. At each time step, the
spatial derivatives are first approximated on the numerical grid thereby
converting the system of PDEs (\ref{linearized-system}\textendash \ref{shock-evolution-equation})
to a system of ODEs:
\begin{align}
\frac{dz'_{i}}{dt} & =-\widehat{L}(\bar{z},z'),\quad i=1,\dots,N-1,\label{eq:numalg:gov}\\
\frac{dM'}{dt} & =\widehat{s}\left(\bar{z},z'\right),\label{eq:numalg:sheveq}
\end{align}
where $\widehat{L}(\bar{z},z',z'_{x})$ and $\widehat{s}(\bar{z},z')$
are approximations of $L(\bar{z},z',z'_{x})=A(\bar{z})z'_{x}+B(\bar{z})z'-\frac{d\bar{z}}{dx}\psi'$
from \eqref{linearized-system} and $s(\bar{z},z',z'_{x})$ from \eqref{shock-evolution-equation},
respectively. In these expressions, the spatial derivatives are evaluated
using finite-difference formulas as given below, and then the system
is evolved in time over one time step $\DeltaT$.

The spatial derivatives are calculated in the following four steps.

1. Approximation at points $i=0,\dots,N-3$. We compute left- and
right-biased approximations of $z'_{x}$ based on the upwind method
of fifth order:
\begin{align}
\left.z_{x}^{-}\right|_{i} & =\frac{-2z_{i-3}+15u_{i-2}-60z_{i-1}+20z_{i}+30z_{i+1}-3z_{i+2}}{60\DeltaX},\label{eq:numal:upwind5-left}\\
\left.z_{x}^{+}\right|_{i} & =\frac{3z_{i-2}-30z_{i-1}-20z_{i}+60z_{i+1}-15z_{i+2}+2z_{i+3}}{60\DeltaX},\label{eq:numalg:upwind5-right}
\end{align}
where subscripts ``-'' and ``+'' denote left- and right-biased
approximations, respectively.

2. Approximation at $i=N-2$. At this grid point, the Taylor expansion
gives a fifth-order formula
\begin{equation}
\left.z'_{x}\right|_{N-2}=\frac{-2z_{N-5}+15z_{N-4}-60z_{N-3}+20z_{N-2}+30z_{N-1}-3z_{N}}{60\DeltaX}.\label{eq:numalg:z_x-N-2}
\end{equation}

3. Approximation at $i=N-1$. At this point, spatial derivatives are
approximated using the fourth-order formula
\begin{equation}
\left.z'_{x}\right|_{N-1}=\frac{-z_{N-4}+6z_{N-3}-18z_{N-2}+10z_{N-1}+3z_{N}}{12\DeltaX}.\label{eq:numalg:z_x-N-1}
\end{equation}

4. Approximation at $i=N$. At this point, only $u'_{x}$ must be
approximated because it appears in the shock-evolution equation \eqref{eq:numalg:sheveq}.
The approximation formula is
\begin{equation}
\left.u'_{x}\right|_{N}=\frac{-12u_{N-5}+75u_{N-4}-200u_{N-3}+300u_{N-2}-300u_{N-1}+137u_{N}}{60\DeltaX}.\label{eq:numalg:z_x-N}
\end{equation}

Expressions \eqref{eq:numalg:z_x-N-2}\textendash \eqref{eq:numalg:z_x-N}
were first used for detonation simulations in \cite{HenrickAslamPowers2006}
in a shock-fitting algorithm.

Once the spatial derivatives are approximated, $\widehat{L}$ is found
using the global Lax\textendash Friedrichs flux: 
\begin{equation}
\widehat{L}(\bar{z},z')=L\left(\bar{z},z',\frac{z_{x}^{'-}+z_{x}^{'+}}{2}\right)-\alpha\frac{z_{x}^{'+}-z_{x}^{'-}}{2},\label{eq:algorithm:lf-flux}
\end{equation}
where $z_{x}^{'-}=z_{x}^{'+}=z'_{x}$ for grid points $i=\{N-2,N-1,N\}$,
$\alpha$ is the largest eigenvalue of $A\left(\bar{z}\right)$ over
the numerical grid, 
\begin{equation}
\alpha=\max_{i=0,\ldots,N}{\{\bar{u}-\bar{c},\bar{u},\bar{u}+\bar{c}\}|_{x=x_{i}}}.
\end{equation}
We found that a simple evaluation of $L$ was not enough to guarantee
numerical stability of simulations, hence the use of the Lax\textendash Friedrichs
flux in \eqref{eq:algorithm:lf-flux}. After evaluation of $L$, $\widehat{s}(\bar{z},z')$
is found using the approximation of $u_{x}|_{\text{s}}$ given by
\eqref{eq:numalg:z_x-N}.

Once $\widehat{L}(\bar{z},z')$ and $\widehat{s}(\bar{z},z')$ are
evaluated on the whole grid, system (\ref{eq:numalg:gov}\textendash \ref{eq:numalg:sheveq})
is evolved in time using the adaptive-step time integrator DOPRI5
\cite{hairer1993solving}, which is an embedded pair of explicit Runge\textendash Kutta
methods of orders 5 and 4 used to estimate a local error. We set both
absolute and relative tolerances of the time integrator to $10^{-14}$.

Simulation proceeds to the final time $\TFinal$, where typically
$\TFinal=10$ is used. While computing the solution, we record the
time series of the perturbation of detonation velocity $\psi'(t)$
at uniform time intervals $\DeltaT=0.005$. When simulation reaches
$\TFinal$, the solver computes the ratio of $L_{2}$-norms of the
time series for $\TFinal/2\leq t\leq\TFinal$ and $0\leq t\leq\TFinal/2$,
and if this ratio is smaller than three, then $\TFinal$ is increased
to 100, as we found it necessary to record longer time series when
the flow is not sufficiently unstable. The analysis of the resulting
time series of the perturbation of detonation velocity allows us to
extract growth rates and frequencies of unstable modes by a post-processing
method described below in Subsection \ref{subsec:algorithm:postproc}.

We also need to provide initial conditions for $z'$ and $M'$. First,
the perturbation of the detonation velocity by a small value $A_{0}$=$10^{-10}$
is specified. Then, the initial value of $M'$ is computed using $M'=-\rhoA A_{0}$.
The Rankine\textendash Hugoniot conditions are then used to find $\rhoS'$,
$\uS'$, $\pS'$ from $M'$. Subsequently, initial conditions for
perturbations over the whole grid are computed using
\begin{equation}
\rho'_{0}(x)=\frac{\rhoS'}{\rhoSZnd}\rhoZnd(x),\quad u'_{0}(x)=\frac{\uS'}{\uSZnd}\uZnd(x),\quad p'_{0}(x)=\frac{\pS'}{\pSZnd}\pZnd(x),\quad\lambda'_{0}(x)=A_{0}\lambdaZnd(x).\label{eq:algorithm:ic}
\end{equation}
As can be seen from these formulas, the initial conditions for perturbations
are specified as a multiple of the ZND solution. We found that such
an initial condition minimizes the transient behavior in the time
series of $\psi'$ at early times. The beginning of the time series
is cut off before the post-processing in order to eliminate the initial
numerical transients from the signal.

Whenever the time integrator evaluates the right-hand side of system
(\ref{eq:numalg:gov}\textendash \ref{eq:numalg:sheveq}), boundary
conditions are imposed as follows. At the shock (grid point $i=N$),
boundary conditions are specified using \eqref{RH-conditions-linearized}.
At the downstream end, no boundary condition should be specified for
the system, however, boundary treatment is necessary for the finite-difference
computations \cite{laney1998}. This is done by the zeroth-order extrapolation:
\[
z_{i}'=z_{0}'\text{ for }i\in\{-3,-2,-1\},
\]
where negative indices are used for the ghost grid points to the left
of the computational domain. For the fifth-order upwind method (\ref{eq:numal:upwind5-left}\textendash \ref{eq:numalg:upwind5-right})
three ghost points are needed.

The algorithm was verified to be fifth order in space and time either
by measuring the convergence rate to the steady-state solution (in
stable cases) or by a self-convergence test (in unstable cases).

\subsection{DMD analysis of the numerical results\label{subsec:algorithm:postproc}}

In the postprocessing step, our goal is to extract growth rates and
frequencies of unstable modes (discrete stability spectrum) from the
time series of detonation velocity 
\begin{equation}
\psi'(t)=\left[\psi'_{0},\psi'_{1},\dots,\psi'_{n}\right]^{\T},\label{eq:algorithm:postproc:time-series}
\end{equation}
where $n$ is the index of the last time step and values $\psi'_{k}$
are defined at times $t_{k}$, $k=0,\dots,n$ such that $t_{k}-t_{k-1}=\DeltaT=\text{const}$
for all $k=1,\dots,n$.

To process such a time series, an efficient and robust algorithm is
required such that the stability spectrum is computed as accurately
as possible. The DMD algorithm \cite{schmid2010dynamic} is at the
heart of our postprocessing. We note however, that the DMD algorithm
does not necessarily answer the question of what the rank of low-dimensional
approximation should be. Hence, we need a way to determine the best
approximation with the ultimate goal being to find all modes that
can also be found by normal-mode analysis.

The remainder of this section proceeds as follows. First, for completeness,
we present a brief generic description of the DMD algorithm. Then,
we describe how we form input data for the DMD algorithm from the
time series \eqref{eq:algorithm:postproc:time-series}, and explain
how we determine the best low-rank approximation. Finally, we apply
the algorithm to synthetic data to assess its performance.

\subsubsection{Description of the DMD algorithm}

Suppose that we have a time series of the snapshots of the state of
some dynamical system:
\begin{equation}
x=[x_{0},\dots,x_{n}]^{\T},\label{eq:algorithm:postproc:given-time-series}
\end{equation}
where $x_{i}\in\R^{m}$ for all $i=0,\dots,n$, and $x_{i}$ being
a snapshot of the system state. Assume also that snapshots are taken
at uniform time intervals $\DeltaT$. Our purpose is to find an eigendecomposition
of a linear operator $A$ such that
\[
x_{k}=Ax_{k-1}
\]
for all $k=1,\dots,n$ at least in the least-squares sense. The eigenvalues
of $A$ contain the information on the growth/decay rates and frequencies
of oscillations that, when combined, represent the time evolution
of the dynamical system. 

To proceed, we define matrices
\begin{equation}
X=[x_{0},\dots,x_{n-1}]^{\T},\quad Y=[x_{1},\dots,x_{n}]^{\T}\label{eq:algorithm:postproc:input-matrices}
\end{equation}
such that $Y=AX$ and thus $A=YX^{\dagger}$, where $X^{\dagger}$
is a Moore\textendash Penrose pseudoinverse of $X$. Although the
problem is formally solved now, we cannot just do the eigendecomposition
of $A$ because $A\in\R^{m\times m}$ with $m$ generally being large.
Instead, a low-rank approximation matrix $\tilde{A}$ and its eigendecomposition
are found. With given inputs~\eqref{eq:algorithm:postproc:input-matrices},
the DMD algorithm proceeds as follows:
\begin{enumerate}
\item Compute reduced singular value decomposition (SVD) of $X$:
\begin{equation}
X=U\Sigma V^{\T},\label{eq:algorithm:postproc:svd}
\end{equation}
with matrices $U$ and $V$ having orthonormal columns and $\Sigma$
being diagonal with singular values $\sigma_{i}$, $i=1,\dots,\min(m,n)$,
on the diagonal.
\item Truncate $U$, $\Sigma$, and $V$ to rank $r$, which yields matrices
$U_{r}$, $\Sigma_{r}$, and $V_{r}$. Truncation is necessary, for
example, because all singular values $\sigma_{i}$ in $\Sigma$ with
$i>r$ are due to the noise in matrix $X$.
\item Compute auxiliary matrix $\tilde{A}\in\R^{r\times r}$:
\[
\tilde{A}=U_{r}^{\T}AU_{r}=U_{r}^{\T}YX^{\dagger}U_{r}=U_{r}^{\T}YV_{r}\Sigma_{r}^{-1}U_{r}^{\T}U_{r}=U_{r}^{\T}YV_{r}\Sigma_{r}^{-1}.
\]
\item Find eigendecomposition of $\tilde{A}$:
\[
\tilde{A}W=W\Lambda,
\]
where $\Lambda=\diag(\lambda_{1},\dots,\lambda_{r})$ is the matrix
of eigenvalues, $W$ the matrix of eigenvectors.
\item Find truncated eigendecomposition of $A$. All eigenvalues of $\tilde{A}$
are the eigenvalues of $A$ as well (the opposite is not necessarily
true as $\tilde{A}\in\R^{r\times r}$ while $A\in\R^{m\times m}$
with $r\leq m$). Matrix of eigenvectors $\Phi$ of $A$ can be found
using 
\[
\Phi=YV_{r}\Sigma_{r}^{-1}W\Lambda^{-1},
\]
with $\Phi$ being a rectangular matrix containing exact DMD modes
as was introduced in \cite{tu2014dynamic}.
\end{enumerate}
Knowing the eigenvalues $\Lambda$ and eigenvectors $\Phi$ of $A$,
we can reconstruct the time series \eqref{eq:algorithm:postproc:given-time-series}:
\begin{equation}
\widehat{x}_{k}=Ax_{k-1}=A^{k}x_{0}=\Phi\Lambda^{k}b,\label{eq:algorithm:postproc:reconstr}
\end{equation}
where $b=\Phi^{\dagger}x_{0}$ is the vector of the initial amplitudes
of DMD modes. Hence, DMD provides modes which are separated in terms
of frequencies (that is, imaginary parts of $\lambda_{i}$).

To measure the quality of low-rank approximation, the relative error
$e_{\text{DMD}}$ and residual error $e_{\text{resid}}$ are computed:
\begin{equation}
e_{\text{DMD}}=\frac{\norm{\widehat{x}-x}_{2}}{\norm x_{2}},\quad e_{\text{resid}}=\norm{Y-\Phi\Lambda\Phi^{\dagger}X}_{2}.\label{eq:algorithm:postproc:errors}
\end{equation}

We also convert discrete-time eigenvalues $\lambda_{i}$ that imply
stability when they are inside the unit disc in $\mathbb{C}$ to the
continuous-time eigenvalues $\alpha_{i}$ that imply stability when
they are in the left half-plane of $\mathbb{C}$ as they are commonly
used in normal-mode analysis. The conversion formula is
\[
\alpha_{i}=\frac{\log(\lambda_{i})}{\DeltaT},\quad i=1,\dots,r.
\]

\subsubsection{Application of DMD to the time series of detonation velocity}

The DMD as it was introduced in \cite{schmid2010dynamic} assumes
that each snapshot in a time series is a high-dimensional vector coming,
for example, from a solution variable on a numerical grid at a given
time step in simulations or from a large number of flow sensors in
experiments. The time series of perturbation of detonation velocity
$\psi'$ \eqref{eq:algorithm:postproc:time-series} that we record
during our simulations are one-dimensional, that is, $\psi'_{k}\in\R$,
$k=0,\dots,n$. As pointed out in \cite[pp. 401--402]{tu2014dynamic},
one-dimensional time series yields the matrix $\tilde{A}$ which is
$1\times1$, therefore it has only one real eigenvalue, which can
only capture an exponential growth or decay, but not oscillations.
To overcome this rank-deficiency problem, we build an input Hankel
matrix $Z$:
\begin{equation}
Z=\begin{bmatrix}\psi'_{0} & \cdots & \psi'_{n-L}\\
\psi'_{1} & \cdots & \psi'_{n-L+1}\\
\vdots & \vdots & \vdots\\
\psi'_{L-1} & \cdots & \psi'_{n}
\end{bmatrix},\label{eq:algorithm:postproc:input-hankel-matrix}
\end{equation}
which stacks time-shifted values of $\psi'$. This approach is commonly
used in other methods of time-series analysis such as Singular Spectral
Analysis \cite{golyandina2013singular}. Number $L$ in \eqref{eq:algorithm:postproc:input-hankel-matrix}
defines the upper bound on the number of modes that we can extract
from the time series. In this paper, $L=1000$ is used. Then, the
input matrices $X$ and $Y$ \eqref{eq:algorithm:postproc:input-matrices}
are built from $Z$ by excluding the last and first columns of $Z$,
respectively.

As time series of $\psi'$ inevitably contains noise induced by truncation
and rounding errors of the numerical computations, truncation of matrices
$U,$ $\Sigma$, and $V$ from \eqref{eq:algorithm:postproc:svd}
with a target rank $r$ will yield a matrix $\tilde{A}\in\R^{r\times r}$
whose eigenvalues are all physical only if $r$ is known \emph{a priori}.
However, if the rank choice is done by a machine, there is always
a possibility that the value of $r$ will be larger than the rank
of the ideal low-rank approximation, hence, the spurious eigenvalues
will be present in the output of DMD (with even more severe consequences
if, for example, instead of a complex-conjugate pair of eigenvalues
only one real eigenvalue would be found). As we aim to apply DMD in
Section~\ref{sec:results} for parametric studies of detonation stability,
an algorithm is needed for choosing the target rank $r$ automatically
and as accurately as possible. The details of the algorithm that we
use are provided below.

First, the beginning of the time series \eqref{eq:algorithm:postproc:time-series}
must be cut off as it is necessary to exclude determining DMD modes
from nonmodal behavior occurring due to the relaxation of the perturbation
solution from volatile initial conditions \eqref{eq:algorithm:ic}
to a superposition of modes. This cutoff time depends on the length
of the time series. Our simulations run by default to final time $\TFinal=10$,
however they can be prolonged as described in section \ref{subsec:linearized-system}.
If the final time of the time series is larger than 10, then the cutoff
time is 10, otherwise it is 1. Then, the input matrix \eqref{eq:algorithm:postproc:input-hankel-matrix}
is built and separated into $X$ and $Y$. After that, candidates
for a possible rank of $X$ are determined by the following procedure.
We consider all singular values $\sigma_{k}$, $k=1,\dots,K$ of $X$
normalized by $\sigma_{1}$ with $K$ being the smallest index corresponding
to $\sigma_{K+1}<10^{-10}$. Then if $\sigma_{k+1}/\sigma_{k}<0.95$
for some $k$, then $k$ is added to the list of possible ranks as
there is a sufficient gap between $\sigma_{k}$ and $\sigma_{k+1}$,
which indicates that $\sigma_{k+1}$ is a spurious singular value.
For each rank from the list of all possible ranks we determine DMD
decomposition and compute the fit and residual errors \eqref{eq:algorithm:postproc:errors}.
Then we consider two smallest fit errors $e_{\text{DMD},1}$ and $e_{\text{DMD},2}$,
and if $0.5\leq e_{\text{DMD},1}/e_{\text{DMD},2}\leq1$, then we
also take into account corresponding residual errors, so that the
DMD decomposition corresponding to the smallest residual error is
considered the best. Otherwise, the best one corresponds to the decomposition
with the fit error $e_{\text{DMD},1}$. Once the best-rank DMD is
determined, its eigenvalues that have the real part larger than $-1$
and positive imaginary part are saved as an output of the postprocessing
algorithm sorted by the frequency and real part, so that the enumeration
of modes is the same as in the normal-mode analysis.

\subsubsection{Testing the method with synthetic data}

Now we apply the postprocessing algorithm to two synthetic time series
chosen to represent typical unstable behavior of the solutions of
system (\ref{eq:numalg:gov}\textendash \ref{eq:numalg:sheveq}).

\textbf{Example 1.} Time series has a single exponentially growing
and oscillating mode,\textbf{ }
\begin{equation}
y_{1}(t)=A\exp(\alpha_{\text{re}}t)\sin(\alpha_{\text{im}}t),\label{eq:algorithm:postproc:example-1}
\end{equation}
where $\alpha=\alpha_{\text{re}}+i\alpha_{\text{im}}=3+2i,$ $i$
being imaginary unit, $t\in[0;51]$.

\textbf{Example 2.} Time series is a superposition of five exponentially
growing and oscillating modes of different growth rates and frequencies,
\begin{equation}
y_{2}(t)=\sum_{i=0}^{4}A\exp(\alpha_{\text{re},i}t)\sin(\alpha_{\text{im},i}t),\label{eq:algorithm:postproc:example-2}
\end{equation}
where $\alpha\in\{0.7+0.1i,0.8+1.57i,0.6+2.76i,0.5+3.88i,0.01+15.62i\}$,
$t\in[0;21]$. For both time series, the amplitude of each mode is
taken as $A=10^{-10}$, and the time step is $\DeltaT=0.01$. These
time series are intentionally contaminated with a Gaussian noise of
amplitude $10^{-13}$, mean zero, and variance one such that the noise
is proportional to the signal:
\begin{equation}
y=y_{\text{clean}}(1+\text{noise}),
\end{equation}
where $y_{\text{clean}}$ is a noiseless time series (either \eqref{eq:algorithm:postproc:example-1}
or \eqref{eq:algorithm:postproc:example-2}). The added noise is to
mimic the noise due to numerical simulation errors.

To assess the performance of the postprocessing algorithm, we measure
relative errors of extracted growth rates and frequencies. For example,
if one particular mode in a time series has an exact eigenvalue $\alpha$
while the postprocessing algorithm finds a corresponding approximate
eigenvalue $\hat{\alpha}$, then the errors are defined as 
\begin{equation}
e_{\text{re}}=\frac{\abs{\alpha_{\text{re}}-\hat{\alpha}_{\text{re}}}}{\alpha_{\text{re}}},\quad e_{\text{im}}=\frac{\abs{\alpha_{\text{im}}-\hat{\alpha}_{\text{im}}}}{\alpha_{\text{im}}}.\label{eq:algorithm:postproc:errors-example}
\end{equation}

In Table \ref{tab:algorithm:postproc:dmd-synthetic-data}, we show
the obtained modes along with the relative errors. As can be seen,
the algorithm was able to recover all the modes present in the time
series. We find that in general $e_{\text{re}}$ is at least one order
of magnitude larger than $e_{\text{im}}$. Nevertheless, the algorithm
was able to recover in the Example 2 the slowest growing mode (with
growth rate $0.01$), albeit with a larger relative error than for
the other modes, even though the growth rate of this mode is 80 times
smaller that the growth rate of the dominant mode $\alpha=0.8+1.57i$.

\begin{table}
\begin{centering}
\begin{ruledtabular}
\begin{tabular}{lddcc}
\multicolumn{1}{c}{Example} & \multicolumn{1}{r}{$\hat{\alpha}_{\text{re}}$} & \multicolumn{1}{r}{$\hat{\alpha}_{\text{im}}$} & \multicolumn{1}{c}{$e_{\text{re}}$} & \multicolumn{1}{c}{$e_{\text{im}}$}\tabularnewline
\hline 
\multirow{1}{*}{1}  & 3.00  &  2.00   & $1.91 \times 10^{-14}$  & $4.80 \times 10^{-14}$ \\
\hline 
\multirow{5}{*}{2}  & 0.70  &  0.10   & $1.33 \times 10^{-11}$  & $3.51 \times 10^{-12}$ \\
                    & 0.80  &  1.57   & $4.19 \times 10^{-13}$  & $4.21 \times 10^{-14}$ \\
                    & 0.60  &  2.76   & $1.42 \times 10^{-12}$  & $5.87 \times 10^{-13}$ \\
                    & 0.50  &  3.88   & $4.69 \times 10^{-12}$  & $1.04 \times 10^{-12}$ \\
                    & 0.01  & 15.62   & $1.08 \times 10^{-07}$  & $1.16 \times 10^{-10}$ \\
\end{tabular}
\end{ruledtabular}

\par\end{centering}
\caption{Results of applying the postprocessing algorithm to synthetic data
\eqref{eq:algorithm:postproc:example-1} (Example 1) and \eqref{eq:algorithm:postproc:example-2}
(Example 2). Columns 2 and 3 show found growth rates and frequencies,
respectively; columns 4 and 5 show the relative errors of growth rates
and frequencies, respectively, computed using \eqref{eq:algorithm:postproc:errors-example}.\label{tab:algorithm:postproc:dmd-synthetic-data}}
\end{table}

These examples demonstrate that the postprocessing algorithm we use
is able to extract stability spectra from time series consisting of
several modes of very different growth rates and frequencies. That
said, it must be clear that there are also limitations to the procedure:
the quality of the extracted spectrum in general depends on the quality
of the time series at hand (its sampling rate, length, noise level,
etc.).

\begin{figure}
\begin{centering}
\includegraphics{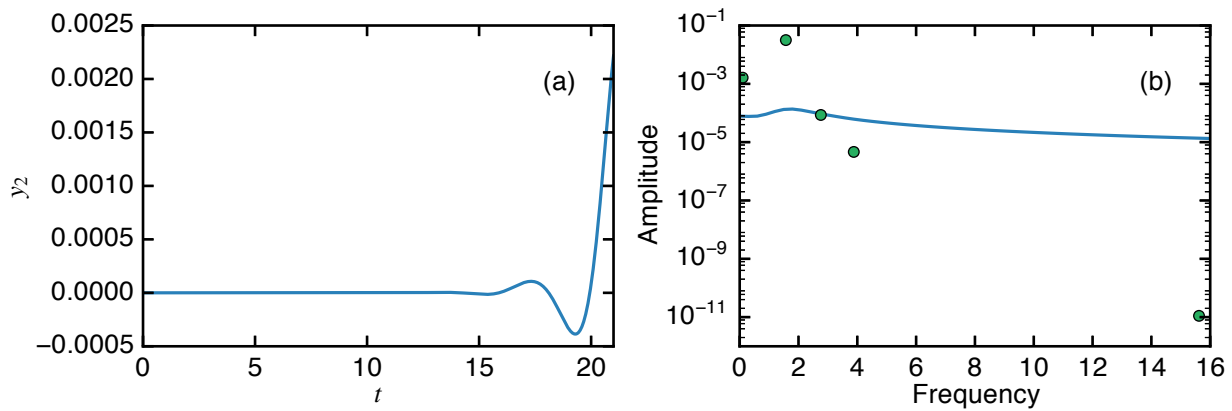}
\par\end{centering}
\caption{\label{fig:Comparison-of-FFT-DMD}Comparison of FFT and DMD algorithms
for analysis of time series \eqref{eq:algorithm:postproc:example-2}:
(a) the time series; (b) the power spectra by FFT (solid line; here
the frequencies are multiplied by $2\pi$) and DMD (markers).}
\end{figure}

As the Fourier transform and power spectra are frequently used in
analysis of oscillatory dynamics, here we contrast this traditional
method with the present approach. Applying the fast Fourier transform
(FFT) to the time series, one can determine approximate values of
frequencies of the modes via the power spectrum and then use nonlinear
least-squares solver to find growth rates of the modes. However note
that for exponentially growing or decaying oscillations, FFT analysis
may be too inaccurate. To illustrate, Figure \ref{fig:Comparison-of-FFT-DMD}
shows the time series \eqref{eq:algorithm:postproc:example-2} and
its power spectra generated via FFT and DMD. While the DMD spectrum
is sparse and is in excellent agreement with the time series, the
FFT spectrum is seen to be essentially flat and dense in the sense
that it is hard to find the particular modes present in the time series. 

The algorithm for extracting growth rates and frequencies of unstable
modes presented in this section requires multiple application of DMD
to find the best low-rank approximation. However, the most expensive
operation in the algorithm (SVD of $X$) is carried out only once.
Our postprocessing code is based on the Intel Math Kernel Library,
which contains optimized routines for numerical linear algebra. The
time that it takes to postprocess our data varies as it is a function
of the length of the time series and the number of possible ranks.
The typical time on a workstation with a Xeon processor is on the
order of a minute. Besides, as we process only one-dimensional time
series \eqref{eq:algorithm:postproc:time-series}, without having
to save the full solution on the numerical grid for each time step,
there is a substantial saving in terms of disk space and corresponding
input-output operations.\selectlanguage{american}%

\selectlanguage{english}%

\section{Stability results\label{sec:results}}

We now apply the numerical algorithm developed above to the detonation
problem with two different reaction-rate mechanisms. In Subsection
\ref{subsec:results:simple}, we consider the conventional detonation
model based on the Arrhenius heat-release rate. This model is widely
used in detonation theory, in particular for stability analysis, since
the pioneering works of Zel'dovich \cite{Zeldovich1940}, von Neumann
\cite{vonNeumann1942}, D\"oring \cite{Doering1943}, and Erpenbeck
\cite{Erpenbeck64,LeeStewart90,short1998cellular,sharpe1997linear}.
We use the results of normal-mode analysis from \cite{LeeStewart90}
and \cite{sharpe1997linear} for verification purposes. In Subsection
\ref{subsec:Two-reaction-model}, we analyze the case of detonation
with one exothermic and one endothermic reactions, and where possible
compare our results with those obtained by normal-mode analysis \cite{sharpe1997linear}.

\subsection{One-step chemistry\label{subsec:results:simple}}

The reaction rate is given by the Arrhenius form, $\omega=k\left(1-\lambda\right)\exp\left(-E/RT\right)$.
The stability of the steady-state self-sustained solutions for this
model depends on the values of $E$, $Q$, and $\gamma$ \cite{Erpenbeck64,LeeStewart90}.
Generally, increasing $E$ has the effect of making the solution more
unstable. The role of $Q$ and $\gamma$ is more complicated. For
example, both very large and very small $Q$ correspond to stable
detonations while there is an instability at intermediate values of
$Q$ if $E$ is large enough. 

\subsubsection{Steady-state solutions, eigenfunctions and eigenvalues}

\begin{figure}
\begin{centering}
\includegraphics{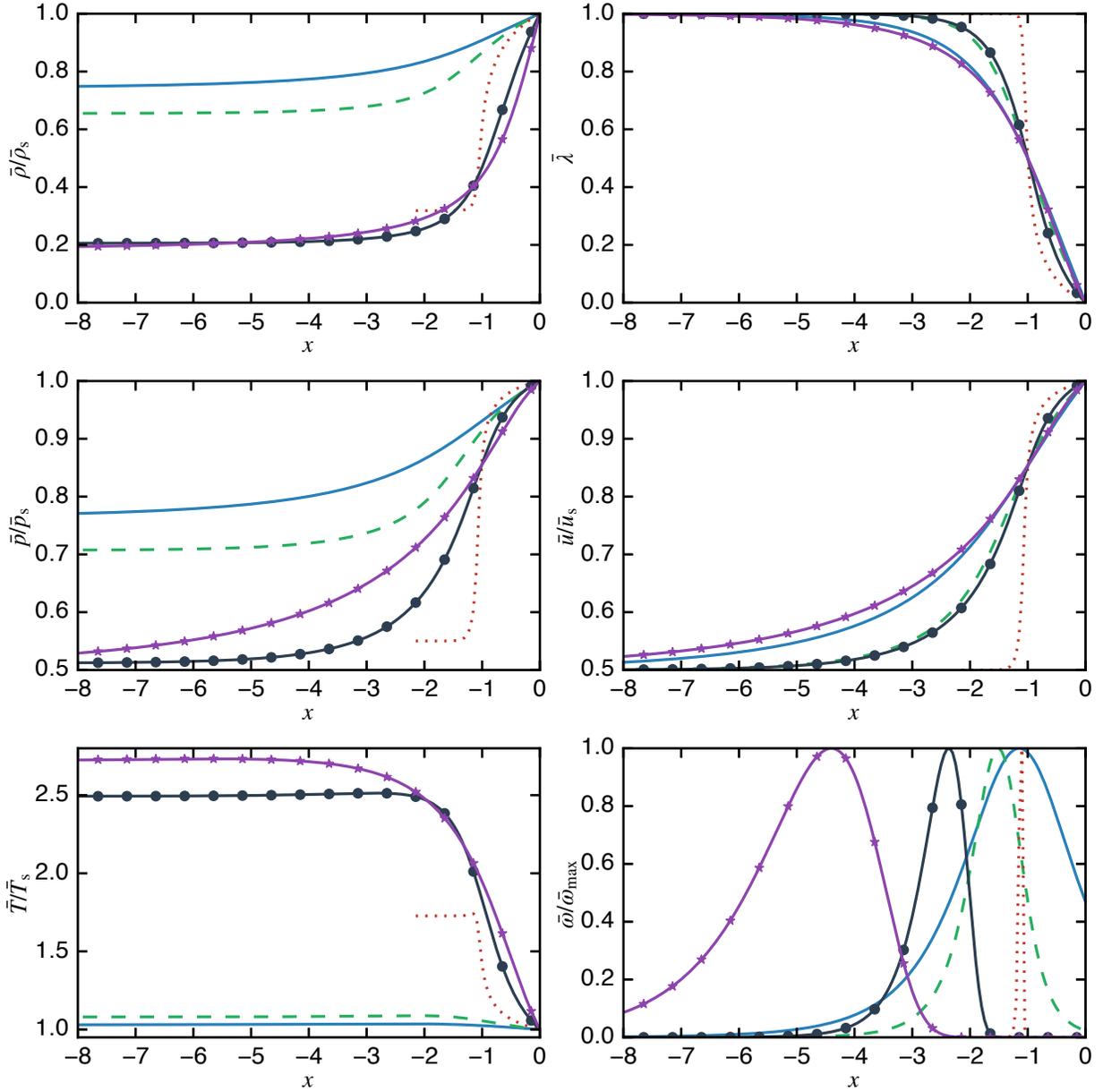}
\par\end{centering}
\caption{The steady solution with fixed $\gamma=1.2,$ $E=30$, and various
values of heat release: $Q=0.5$ - solid, $Q=1.0$ - dashed, $Q=10$
- dotted, $Q=50$ - solid with circles, $Q=100$ - solid with stars.
Density, pressure, velocity in the laboratory frame, and temperature
are normalized by their respective values at the von Neumann (shock)
state; the reaction rate is normalized by its maximum value.\label{fig:results:simple:znd-solutions}}
\end{figure}

The steady-state traveling-wave solutions of (\ref{eq:mass}-\ref{eq:reaction})
are found by setting the time derivatives in the system to zero. In
figure \ref{fig:results:simple:znd-solutions}, we show such ZND solution
profiles computed with fixed parameters $\gamma=1.2$ and $E=30$,
and for $Q=\{0.5,1,10,50,100\}$. For this particular value of $E$,
the case $Q=50$ corresponds to a strongly unstable solution, while
all other values of $Q$ correspond to stable or weakly unstable solutions.
As can be seen from the steady-state profiles, the strongly unstable
case occurs for the ZND solution approaching a square-wave shape with
a spiky heat-release zone, while all the other cases correspond to
relatively smooth profiles. This effect of the steady-state solutions
on their stability is well known. The limiting case of a true square-wave
detonation with instantaneous heat release is known to possess a pathological
spectrum with infinite number of unstable eigenvalues with growth
rates increasing unboundedly with the mode number (or its frequency).
In other words, the square-wave stability problem is ill-posed. However,
with a finite-rate chemistry, the problem remains well-posed even
though the sharper the heat-release rate the larger the number of
unstable modes and the larger their growth rates and frequencies.
Computing the latter situations becomes a challenging problem.

Now we analyze the time series of the perturbation of detonation
velocity, $\psi'$, for two cases that differ by the complexity of
their unstable spectra. The parameters for these cases are $\gamma=1.2$,
$Q=50$, and $E=35$ or $E=40$. Linear stability analysis \cite{sharpe1997linear}
predicts that both of these cases have several unstable modes. In
addition, the fundamental mode for the second case is purely exponential
and has two branches. 

\begin{figure}
\begin{centering}
\includegraphics{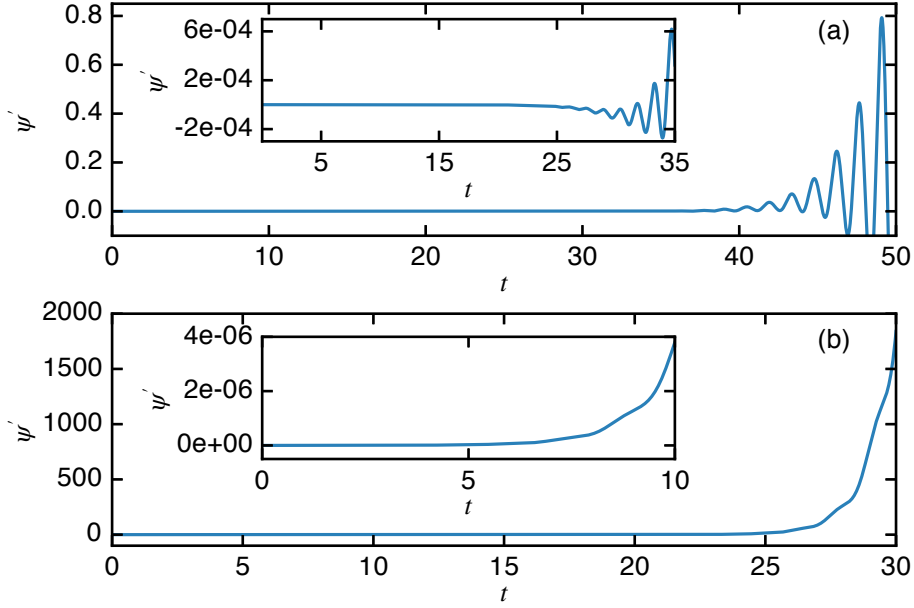}
\par\end{centering}
\caption{Computed time series of the perturbation of detonation velocity $\psi'$
for $\gamma=1.2$, $Q=50$, and (a) $E=35$ or (b) $E=40$. Insets
show the evolution of $\psi'$ at early time of simulation.\label{fig:51-2017-04-09:raw-data-examples}}
\end{figure}

Simulations are conducted with $\TFinal=50$ for $E=35$ and $\TFinal=40$
for $E=40$. Figure \ref{fig:51-2017-04-09:raw-data-examples} displays
the computed time series, with insets showing the solution at early
times. The plots suggest that for $E=35$ all modes are oscillatory,
while for $E=40$ an exponential nonoscillatory growth is present.
Table \ref{tab:51-2017-04-09:raw-data-modes} shows the unstable spectra
for both of these cases obtained with $\Nhrz=1280$ as well as relative
errors calculated by comparing the modes found with $\Nhrz=640$ and
$\Nhrz=1280$. For the case $E=40$, the number of modes found is
the same as shown on figure 6 of \cite{sharpe1997linear}. The fundamental
mode is purely exponential and is split in two branches in agreement
with \cite{sharpe1997linear} where the author notes that the split
occurs for $E\geq35.86$. The growth rate of the upper branch of the
fundamental mode is 1.008 in our calculations while it is 1.021 in
\cite{sharpe1997linear} (extracted from figure 6 of \cite{sharpe1997linear}
and rescaled to the Erpenbeck scales).
\begin{table}
\begin{centering}
\begin{ruledtabular}
\begin{tabular}{l l D{.}{.}{5} c D{.}{.}{5} c}
$E$ &
{$i$} &
\multicolumn{1}{c}{$\alpha_{\text{re}}$} &
{$e_{\text{re}}$} &
\multicolumn{1}{c}{$\alpha_{\text{im}}$} &
{$e_{\text{im}}$} \\
\hline
\multirow{2}{*}{35}
 & 0 & 0.42323 & $3 \times 10^{-11}$ & 0.17957 & $2 \times 10^{-10}$ \\
 & 1 & 0.50528 & $1 \times 10^{-11}$ & 4.36602 & $1 \times 10^{-13}$ \\
\hline
\multirow{5}{*}{40}
 & 0.0 & 0.18851 & $3 \times 10^{-03}$ &  0.00000 & $2 \times 10^{-16}$ \\
 & 0.1 & 1.00847 & $5 \times 10^{-12}$ &  0.00000 & $2 \times 10^{-16}$ \\
 & 1   & 1.01845 & $2 \times 10^{-12}$ &  4.30457 & $3 \times 10^{-13}$ \\
 & 2   & 0.67850 & $8 \times 10^{-09}$ &  7.90388 & $3 \times 10^{-10}$ \\
 & 3   & 0.18504 & $4 \times 10^{-03}$ & 11.43653 & $6 \times 10^{-05}$ \\
\end{tabular}
\end{ruledtabular}

\par\end{centering}
\caption{Unstable spectra for $\gamma=1.2$, $Q=50$, and $E=35$ or $E=40$
obtained from simulations with $\protect\Nhrz=1280$. Here, $i$ is
the mode number; $\alpha_{\text{re}}$ and $\alpha_{\text{im}}$ are
the growth rates and frequencies, respectively; $e_{\text{re}}$ and
$e_{\text{im}}$ are the corresponding relative errors computed from
the results with $\protect\Nhrz=640$ and $\protect\Nhrz=1280$. \label{tab:51-2017-04-09:raw-data-modes}}
\end{table}

\begin{figure}
\begin{centering}
\includegraphics{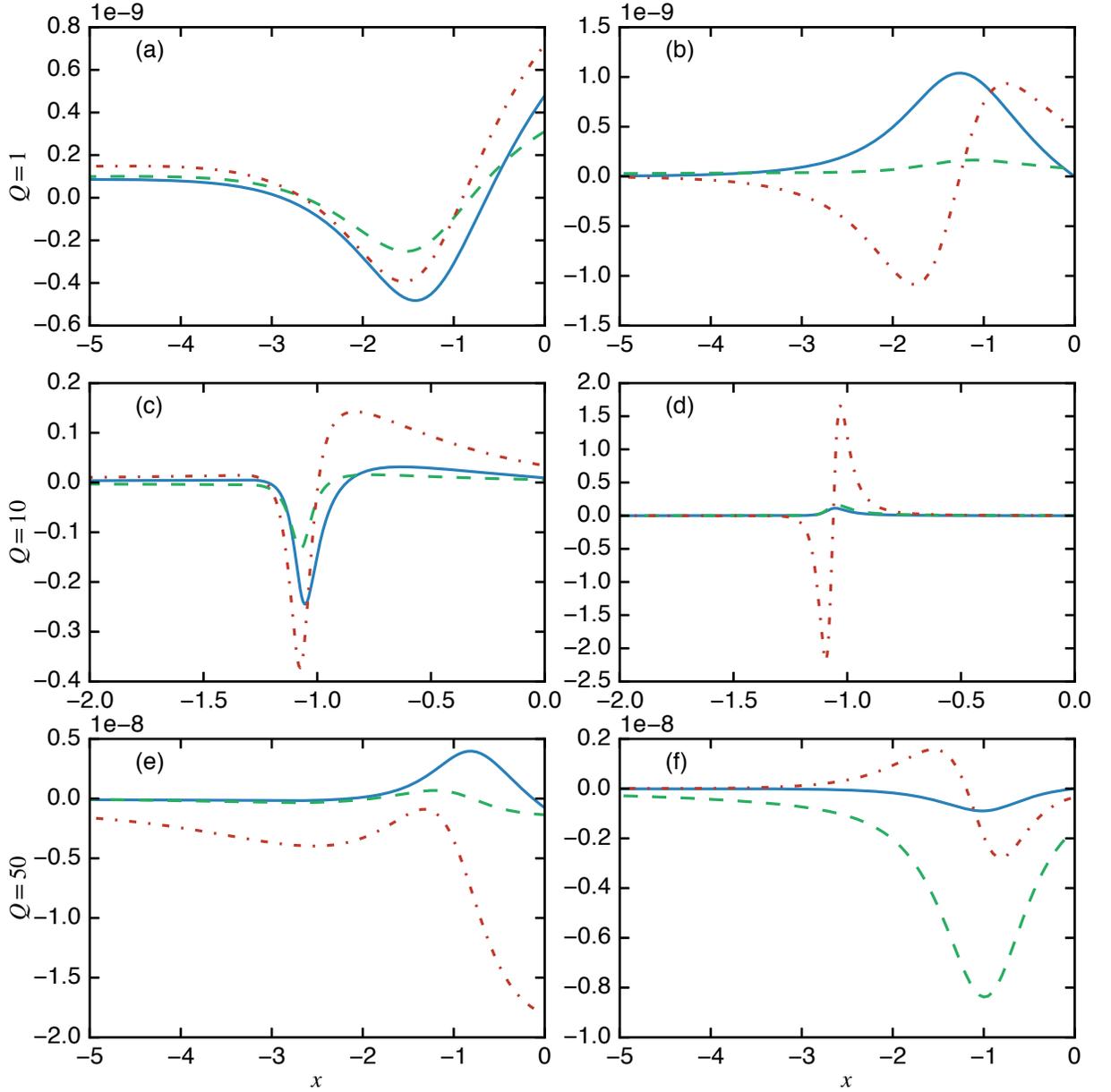}
\par\end{centering}
\caption{Perturbation profiles versus $x$ for $E=30$ and (a-b) $Q=1$, (c-d)
$Q=10$, (e-f) $Q=50$. Left column: $\rho'$ (solid), $u'$ (dashed),
$p'$ (dash-dotted); right column: $\lambda'$ (solid), $T'$ (dashed),
$\omega'$ (dash-dotted). These solutions are plotted at a particular
time, hence the scales are arbitrary. }
\label{fig:results:pert-profiles-vs-x} 
\end{figure}

During the integration of the linearized system (\ref{linearized-system}\textendash \ref{shock-evolution-equation})
we obtain the spatial profiles of the perturbations that are superpositions
of the eigenfunctions of the problem. Such profiles are shown in Figure~\ref{fig:results:pert-profiles-vs-x}
at $E=30$ and at three values of $Q$, two of which are near the
neutral curve and one somewhere in the middle of the unstable domain.
These are $Q=1$ (stable), $Q=10$ (strongly unstable), and $Q=50$
(weakly unstable).

\subsubsection{Comparison with normal-mode results}

Here, we show a quantitative comparison with the normal-mode results
found in the literature. As other researchers provide explicit values
for unstable spectra for parameters $\gamma=1.2$, $Q=50$, we take
these parameters as well and vary activation energy $E$. Table \ref{tab:results:comparison-with-normal-modes}
shows the results of the comparison. Our results are in agreement
with the normal-mode results to at least two significant digits for
all considered cases. The normal-mode results are borrowed from the
work of different researchers: \cite{sharpe1997linear} for $E=25.26$
and $E=31.05$, from \cite{HenrickAslamPowers2006} for $E=26$, and
from \cite{LeeStewart90} for $E=50$. These researchers all used
different scales, which are converted here to the same scales for
comparison. We choose to convert to the Erpenbeck scales used in the
present work. The conversion factors are as follows. Conversion factor
from the scales of \cite{sharpe1997linear} to the Erpenbeck scales
is nondimensional $\bar{D}$ in the Erpenbeck scales, which is 6.809475
for $\gamma=1.2$ and $Q=50$. In \cite{LeeStewart90}, the stability
spectrum for $E=50$ is given in scales of \cite{AbouseifToong82},
and the following conversion factor should be used:
\[
\frac{k}{\int_{0}^{1/2}\left(1-\lambdaZnd\right)^{-1}\exp\left(\rhoZnd E/\pZnd\right)\dd\lambdaZnd},
\]
where $k$ is the nondimensional Arrhenius rate constant computed
using \eqref{eq:k}, and the integral in the denominator is the nondimensional
Arrhenius rate constant in the scales of \cite{AbouseifToong82}.
For $\gamma=1.2$, $Q=50$, and $E=50$, the numerical value of this
factor is 0.938749. Conversion of results of \cite{HenrickAslamPowers2006}
is not needed because they provide the stability spectrum in the same
scales as ours.

To obtain the results shown in Table \ref{tab:results:comparison-with-normal-modes},
we run the simulations to final time $\TFinal=30$ for $E=25.26$
and $E=26$ and to $\TFinal=10$ for $E=31.05$ and $E=50$. The simulations
are conducted with resolutions $\Nhrz=20,\dots,1280$ with refinement
factor 2 to verify the convergence of the obtained stability spectra,
which is observed as expected. Table \ref{tab:results:comparison-with-normal-modes}
shows the spectra obtained with $\Nhrz=1280$. The errors on growth
rates and frequencies are also shown in Table \ref{tab:results:comparison-with-normal-modes}.
They are computed using:
\[
e_{\text{re}}=\abs{\frac{\alpha_{\text{re},1}-\alpha_{\text{re},2}}{\alpha_{\text{re},2}}},\quad e_{\text{im}}=\abs{\frac{\alpha_{\text{im},1}-\alpha_{\text{im},2}}{\alpha_{\text{im},2}}},
\]
where subscripts 1 and 2 denote quantities obtained with $\Nhrz=640$
and $\Nhrz=1280$, respectively.

\begin{table}
\begin{centering}
\begin{ruledtabular}
\begin{tabular}{l l D{.}{.}{5} c D{.}{.}{5} c D{.}{.}{4} D{.}{.}{5}}
  &
  &
\multicolumn{4}{c}{Present work} &
\multicolumn{2}{c}{Normal-mode analysis} \\
\cline{3-6} \cline{7-8}
\multicolumn{1}{c}{$E$} &
\multicolumn{1}{c}{$i$} &
\multicolumn{1}{c}{$\alpha_{\text{re}}$} &
\multicolumn{1}{c}{$e_{\text{re}}$} &
\multicolumn{1}{c}{$\alpha_{\text{im}}$} &
\multicolumn{1}{c}{$e_{\text{im}}$} &
\multicolumn{1}{c}{$\alpha_{\text{re}}$} &
\multicolumn{1}{c}{$\alpha_{\text{im}}$} \\
\hline
\multirow{1}{*}{$25.26^{*}$}
& 0   & -0.00017 & $1 \times 10^{-04}$ &  0.53048 & $6 \times 10^{-08}$ & 0.000   & 0.530 \\
\hline
\multirow{1}{*}{$26^{**}$}
& 0   &  0.03709 & $5 \times 10^{-09}$ &  0.52215 & $5 \times 10^{-11}$ & 0.0371 & 0.52215 \\
\hline
\multirow{2}{*}{$31.05^{*}$}
& 0   &  0.26756 & $1 \times 10^{-10}$ &  0.40280 & $3 \times 10^{-10}$ & \multicolumn{1}{c}{N/A} & \multicolumn{1}{c}{N/A} \\
& 1   & -0.00060 & $2 \times 10^{-08}$ &  4.37774 & $1 \times 10^{-12}$ & 0.00    & 4.38 \\
\hline
\multirow{12}{*}{$50^{***}$}
& 0.0 &  0.09365 & $3 \times 10^{-07}$ &  0.00000 & $2 \times 10^{-16}$ & 0.084   & 0.000 \\
& 0.1 &  1.74458 & $2 \times 10^{-11}$ &  0.00000 & $2 \times 10^{-16}$ & 1.743   & 0.000 \\
& 1   &  1.76536 & $6 \times 10^{-11}$ &  4.10817 & $1 \times 10^{-12}$ & 1.764   & 4.104 \\
& 2   &  1.77399 & $6 \times 10^{-12}$ &  7.83005 & $2 \times 10^{-11}$ & 1.772   & 7.823 \\
& 3   &  1.67745 & $1 \times 10^{-09}$ & 11.42106 & $1 \times 10^{-10}$ & 1.676   & 11.42 \\
& 4   &  1.53541 & $1 \times 10^{-08}$ & 14.99313 & $4 \times 10^{-10}$ & 1.534   & 14.98 \\
& 5   &  1.34788 & $4 \times 10^{-08}$ & 18.55077 & $6 \times 10^{-10}$ & 1.346   & 18.53 \\
& 6   &  1.14096 & $1 \times 10^{-07}$ & 22.10524 & $7 \times 10^{-10}$ & 1.140   & 22.09 \\
& 7   &  0.91468 & $5 \times 10^{-07}$ & 25.65854 & $2 \times 10^{-10}$ & 0.913   & 25.64 \\
& 8   &  0.67788 & $1 \times 10^{-06}$ & 29.21139 & $2 \times 10^{-09}$ & 0.677   & 29.19 \\
& 9   &  0.43092 & $4 \times 10^{-06}$ & 32.76403 & $5 \times 10^{-09}$ & 0.431   & 32.73 \\
& 10  &  0.17764 & $2 \times 10^{-05}$ & 36.31692 & $1 \times 10^{-08}$ & 0.177   & 36.28 \\
\end{tabular}
\end{ruledtabular}

\par\end{centering}
\caption{Comparison of the unstable spectra between the present work and the
normal-mode results for $\gamma=1.2$ and $Q=50$. Sources for the
normal-mode results: $^{*}$\cite[p. 2617]{sharpe1997linear}; $^{**}$\cite[p. 321]{HenrickAslamPowers2006};
$^{***}$\cite[p. 127]{LeeStewart90} except mode 0.0 which is extracted
from \cite[figure 6]{short1998cellular}. Here, $i$ is the mode number.
For $E=50$ the fundamental mode has two branches, which we denote
with $0.0$ and $0.1$. Eigenvalues $\alpha$ are in the scales used
in the present work. Notice that the errors are computed using grid
refinement.\label{tab:results:comparison-with-normal-modes}}
\end{table}

\subsubsection{Neutral stability curves}

Next, we use the present method to generate neutral stability curves
in $E$\textendash $Q$ plane for various $\gamma$. The corresponding
algorithm is as follows. Assuming that instability monotonically increases
as $E$ increases, the idea of bisection is used to find a critical
value of $E$ for each given $Q$ such that the growth rate of a given
mode is close to zero with tolerance $10^{-4}$. Initial interval
of search for $E$ is taken $[10;140]$ for all results shown in this
section. Grid resolution used here is $\Nhrz=40$. 
\begin{figure}
\centering\includegraphics{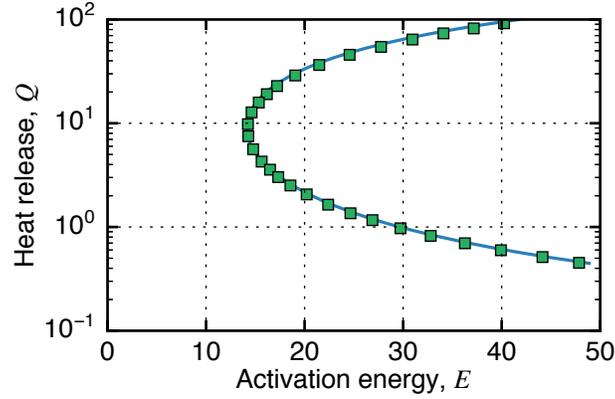}\caption{The neutral stability boundary computed with the present method (solid
line) and its comparison with the normal-mode result of Lee and Stewart
\cite[figure 7]{LeeStewart90} (square markers).}
\label{fig:LS-comparison} 
\end{figure}

\begin{figure}
\centering \includegraphics{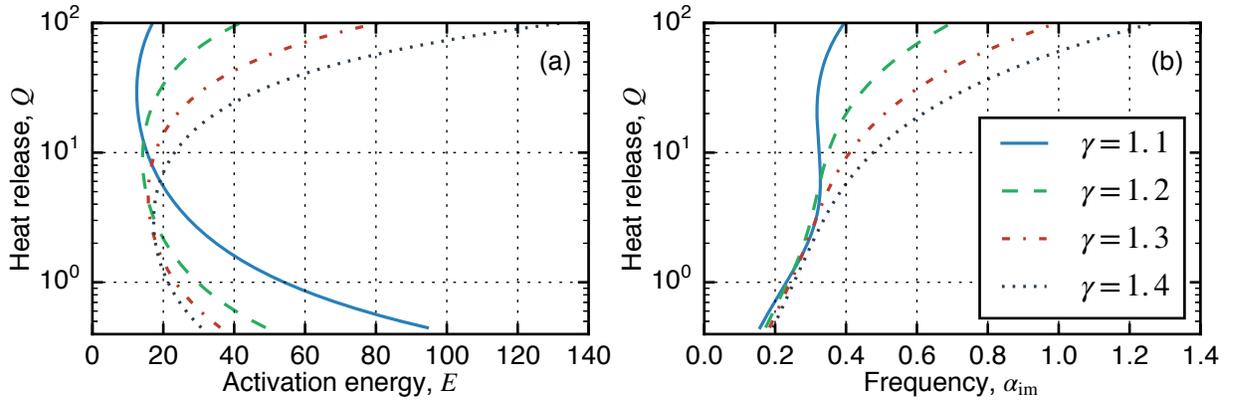} \caption{Neutral stability at various $\gamma$: (a) neutral stability boundary
in $E$\textendash $Q$ plane, (b) frequency of oscillation along
the neutral stability boundary.\label{fig:neutral-boundary-var-gamma}}
\end{figure}

Figure \ref{fig:LS-comparison} displays the comparison between our
results and those of \cite{LeeStewart90} for the fundamental mode
and $\gamma=1.2$. It can be seen that there is a good agreement between
the results. In Figure \ref{fig:neutral-boundary-var-gamma}(a), the
neutral stability curves of mode 0 in $E$\textendash $Q$ plane are
shown for various $\gamma$, while in Figure \ref{fig:neutral-boundary-var-gamma}(b),
the frequency of neutral oscillation is shown. The smaller values
of $\gamma$ are seen to extend the range of unstable $E$ to smaller
$E$ when $Q$ is larger than about $4$, and to reduce the unstable
range towards larger values of $E$ at $Q\lessapprox4$. At the same
time, the frequency of neutral oscillation is seen to decrease substantially
with decreasing $\gamma$ at large $Q$, while it is essentially independent
of $\gamma$ at small values of $Q$. However, for small $Q$ frequency
exhibits nonmonotonic behavior for all $\gamma$ except $\gamma=1.4$.
 
\begin{figure}
\begin{minipage}[t][1\totalheight][b]{0.49\textwidth}%
\begin{center}
\includegraphics{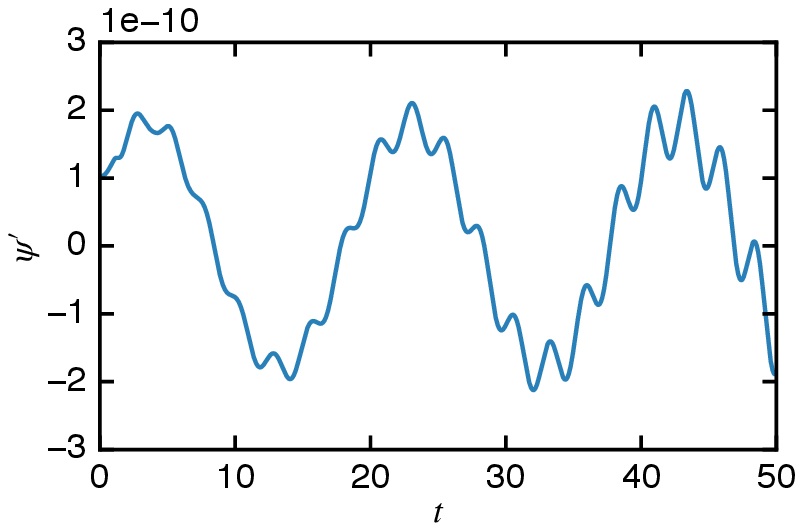}\caption{Example time series of $\psi'$ for $\gamma=1.1$, in which mode 0
is neutrally stable while mode 1 is unstable.\label{fig:time-series-for-gamma=00003D1.1}}
\par\end{center}%
\end{minipage}\hfill{}%
\begin{minipage}[t][1\totalheight][b]{0.49\textwidth}%
\begin{center}
\includegraphics{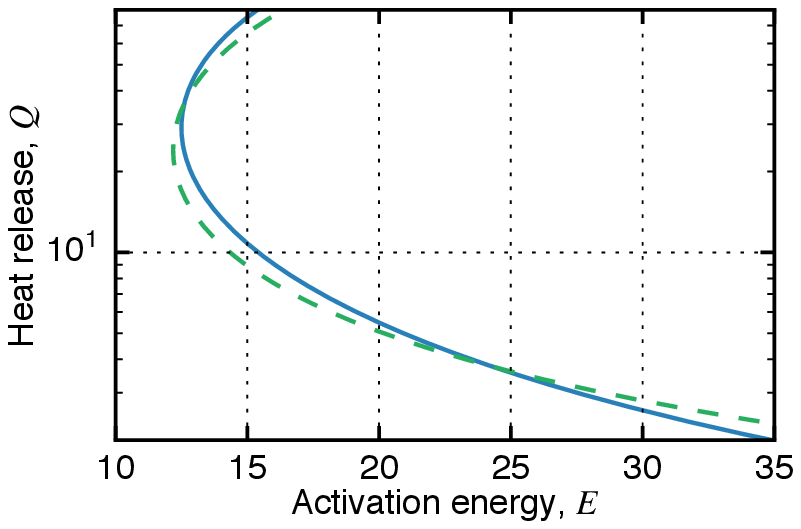}\caption{Neutral stability curves for $\gamma=1.1$ for modes 0 (solid line)
and 1 (dashed line). \label{fig:neutral-curves-gamma=00003D1.1}}
\par\end{center}%
\end{minipage}
\end{figure}

For $\gamma=1.2$, $1.3$, and $1.4$ neutral stability curves shown
in Figure \ref{fig:neutral-boundary-var-gamma}(a) represent stability
boundaries for corresponding $\gamma$, which separate stable ZND
solutions from unstable ones. However, the case $\gamma=1.1$ is more
complicated as for $3.7\lessapprox Q\lessapprox36$ mode 1 is more
unstable than the fundamental mode. A typical time series of the perturbation
of detonation velocity $\psi'$ for such situation is shown in Figure
\ref{fig:time-series-for-gamma=00003D1.1} for $Q=30.473$ and $E=12.503$,
which corresponds to the mode 0 being neutrally stable (its growth
rate is about $-10^{-6}$) while mode 1 grows with a rate about $0.03$.
Figure \ref{fig:neutral-curves-gamma=00003D1.1} shows neutral stability
curves for both mode 0 and 1 for $\gamma=1.1$ on the same plot, where
it can be clearly seen that the neutral stability boundary is determined
by a composition of the neutral stability curves of these two modes.
This intersection of neutral curves at $\gamma=1.1$ was not mentioned
before in the literature, to the best of our knowledge. 

\subsection{Two-reaction model \label{subsec:Two-reaction-model}}

The model with two reactions represents an example of complex chemistry
that brings in the following additional complications that are absent
in the one-step model. First, we now have two traveling-wave solutions
for the same set of parameters of the problem, and therefore we must
investigate the stability of both. Second, the solution has an embedded
sonic point, that is the sonic condition is reached at a finite distance
from the lead shock. This divides the post-shock flow into two regions
that play distinct roles in the detonation dynamics. Following prior
research on the role of sonic points in unsteady detonation dynamics
\cite{kasimov2004dynamics,StewartKasimovSIAP05}, we expect that the
dynamics is affected only by the solution between the shock and the
sonic point. How exactly this idea plays out in the linear stability
calculations is not obvious, and this will be explored in this section.

\subsubsection{Steady-state solutions and perturbations}

Now the model has several additional parameters coming from the more
complex chemistry. Comprehensive investigation of the entire parameter
space would be a difficult task. Instead, we focus here on particular
questions that have certain physical interest. Namely, we investigate
how the level of endothermicity affects stability, and in addition,
we compare stability behavior of the two steady-state solutions that
co-exist at the same parameter set. 

\begin{figure}
\begin{centering}
\includegraphics{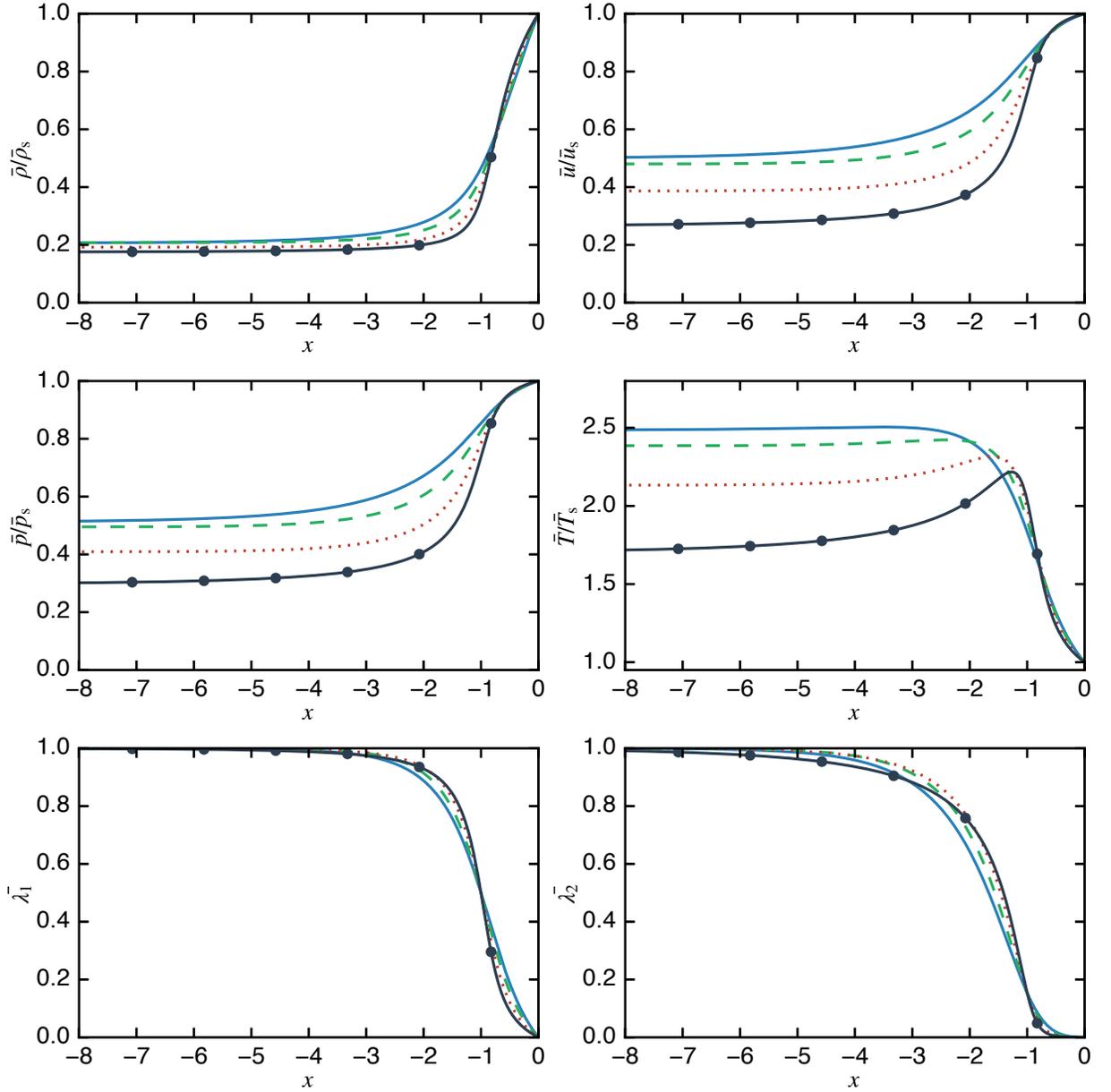}
\par\end{centering}
\caption{ZND subsonic-supersonic profiles of density $\bar{\rho}$, flow velocity
$\bar{u}$, pressure $\bar{p}$, temperature $\bar{T}$, and progress
variables $\bar{\lambda}_{1}$ and $\bar{\lambda}_{2}$ for $\gamma=1.2$,
$Q_{1}=50$, $E=25$, and different values of $Q_{2}$. Solid line
corresponds to $Q_{2}=-1$; dashed line to $Q_{2}=-10$; dotted line
to $Q_{2}=-20$; solid line with circle markers to $Q_{2}=-30$. Density,
flow velocity, pressure, and temperature are normalized by their values
on the shock.\label{fig:eigval-znd-solutions-sub-super}}
\end{figure}

First, we study the behavior of ZND solutions when $Q_{2}$ is varied
while all other parameters are kept fixed. Figure \ref{fig:eigval-znd-solutions-sub-super}
shows profiles of ZND variables $\rhoZnd$, $\uZnd$, $\pZnd$, $\TZnd$,
$\bar{\lambda}_{1}$, and $\bar{\lambda}_{2}$ for subsonic-supersonic
case with $\gamma=1.2$, $Q_{1}=50$ and $E=25$ for several values
of $Q_{2}$, obtained with $\Nhrz=40$. As can be seen from the figure,
the lower the values of $Q_{2}$, the larger the gradients of the
profiles become. The maximum temperature decreases with increasing
magnitude of $Q_{2}$, as expected. 

\begin{figure}
\begin{centering}
\includegraphics{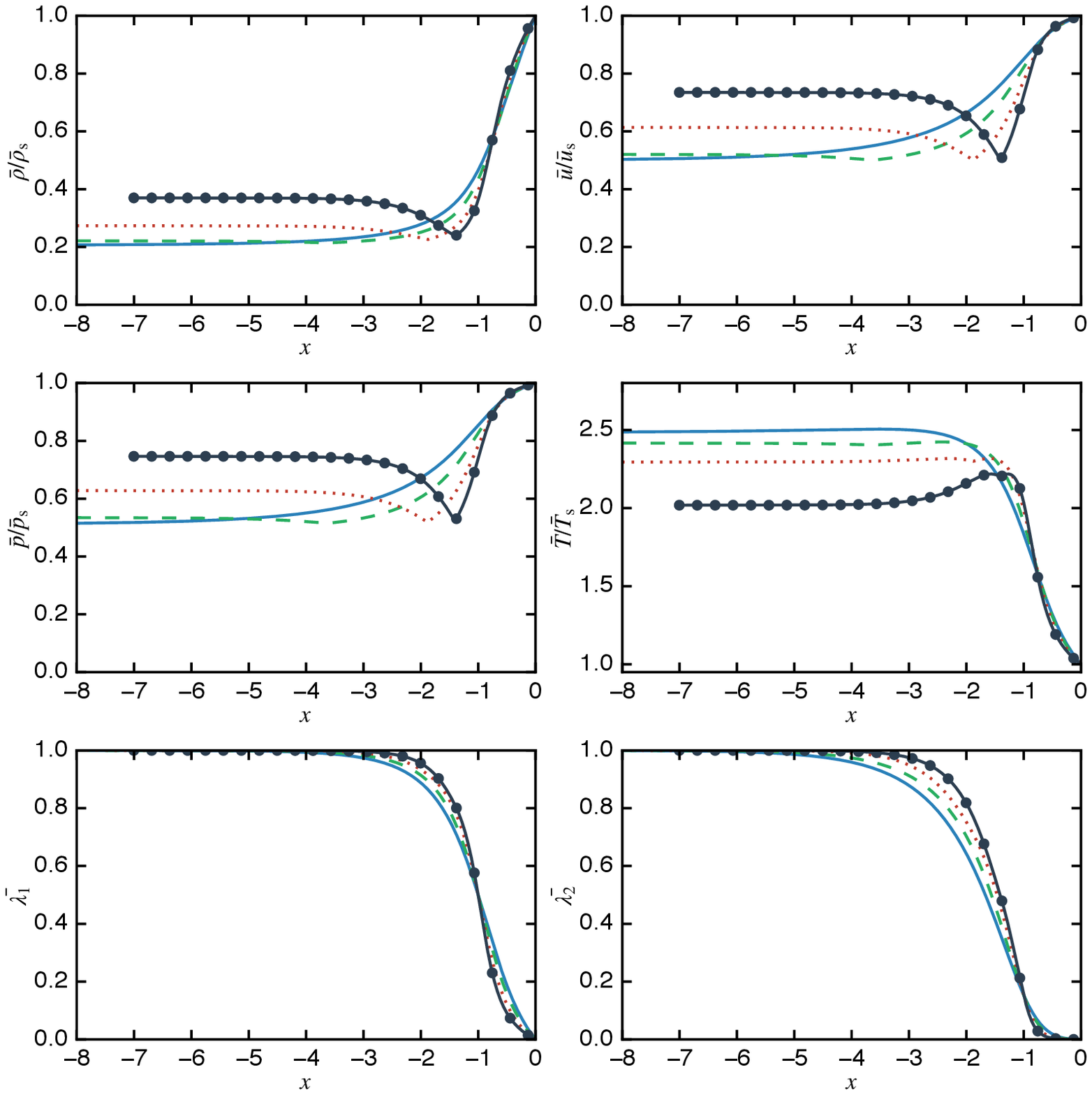}
\par\end{centering}
\caption{ZND subsonic-subsonic profiles of density $\bar{\rho}$, flow velocity
$\bar{u}$, pressure $\bar{p}$, temperature $\bar{T}$, and progress
variables $\bar{\lambda}_{1}$ and $\bar{\lambda}_{2}$ for different
values of $Q_{2}$. Solid line corresponds to $Q_{2}=-1$; dashed
line to $Q_{2}=-10$; dotted line to $Q_{2}=-20$; solid line with
circle markers to $Q_{2}=-30$. Density, flow velocity, pressure,
and temperature are normalized by their values on the shock.\label{fig:fig:eigval-znd-solutions-sub-sub}}
\end{figure}

Figure \ref{fig:fig:eigval-znd-solutions-sub-sub} displays the same
results as in Figure \ref{fig:eigval-znd-solutions-sub-super}, but
for subsonic-subsonic solution, which is obtained now with $\Nhrz=640$
to accurately resolve the weak discontinuity in the solution. It can
be seen that as the magnitude of $Q_{2}$ increases, and the sonic
point moves toward the shock, the kink in the profiles of $\rhoZnd$,
$\uZnd$ and $\pZnd$ becomes more pronounced. 

\begin{figure}
\begin{centering}
\includegraphics{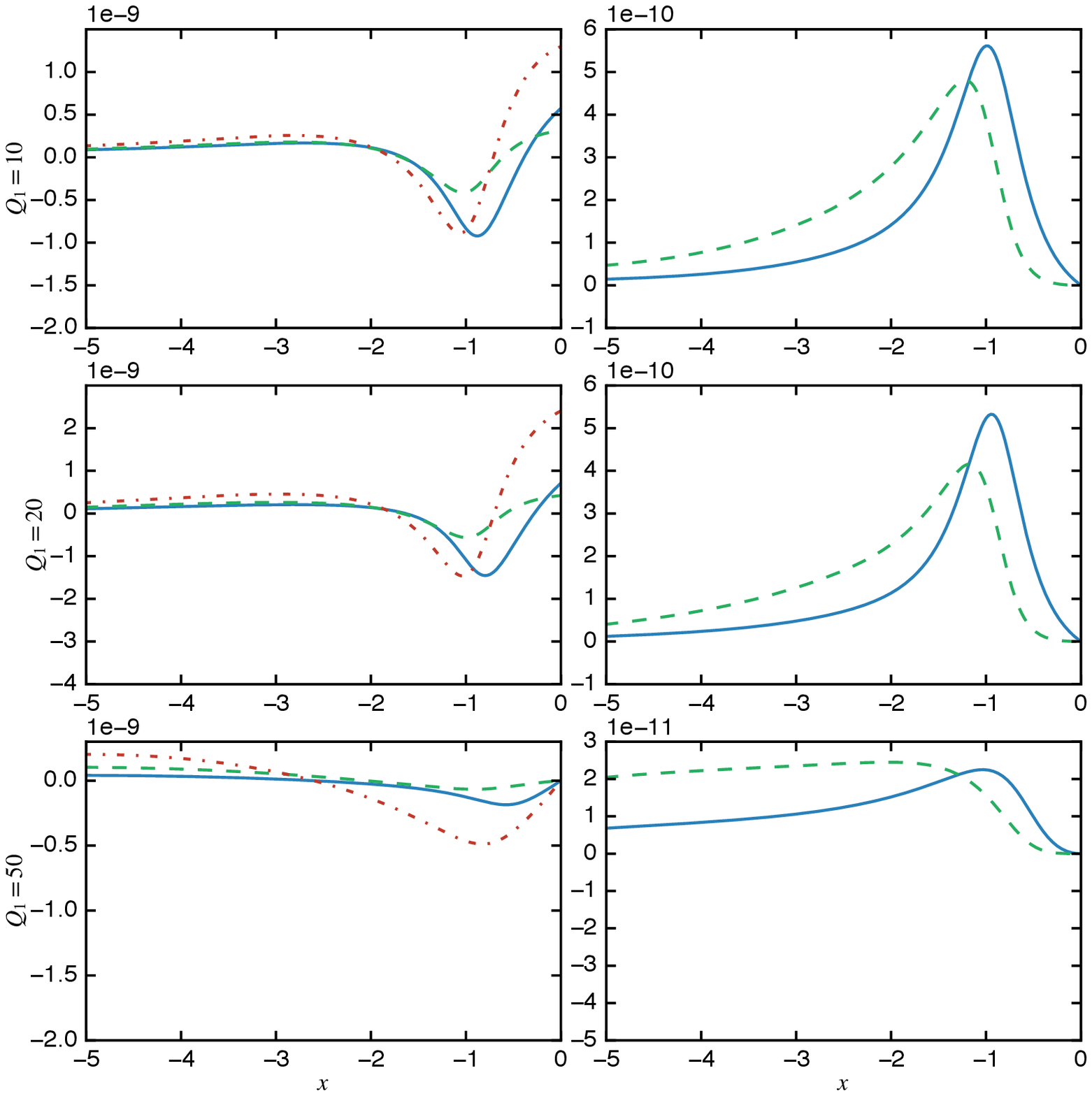}
\par\end{centering}
\caption{Perturbation profiles for the subsonic-supersonic case for $\gamma=1.2$,
$E=15$, $Q_{1}\in\{10,20,50\}$ with $Q_{2}=-0.75Q_{1}$. On the
left: solid line $\rho'$, dashed line $u'$, dash-dotted line $p'$.
On the right: solid line $\lambda_{1}'$, dashed line $\lambda_{2}'$.
Perturbations are not scaled and shown as they are at the simulation
time $\protect\TFinal=4$. Notice, that the left limits of $x$-axes
are taken to be 5 for clarity and are much smaller than the reaction
lengths.\label{fig:eigval-pert}}
\end{figure}

\begin{figure}
\begin{centering}
\includegraphics{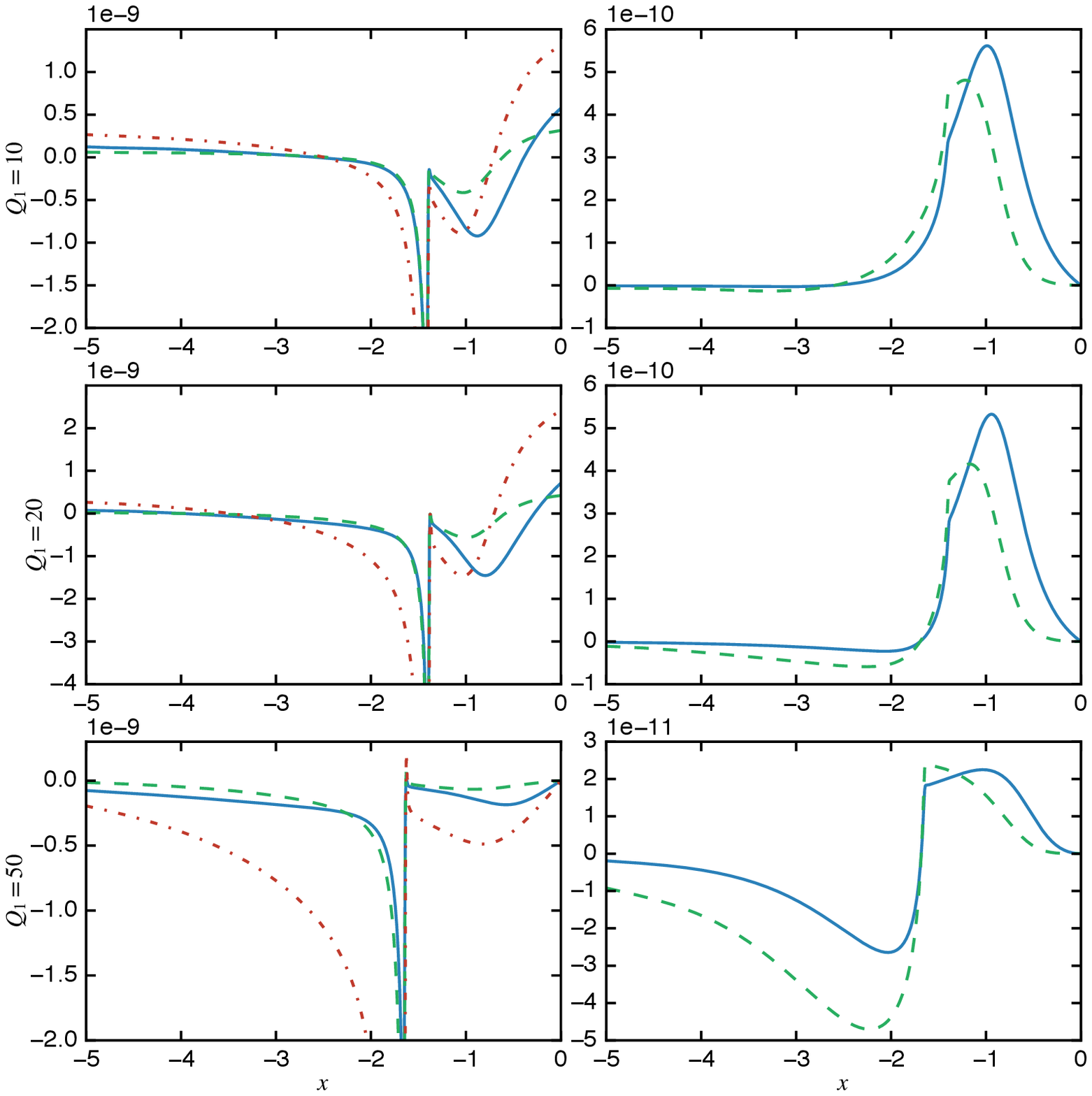}
\par\end{centering}
\caption{Perturbation profiles for the subsonic-subsonic case for $\gamma=1.2$,
$E=15$, $Q_{1}\in\{10,20,50\}$ with $Q_{2}=-0.75Q_{1}$. On the
left: solid line $\rho'$, dashed line $u'$, dash-dotted line $p'$.
On the right: solid line $\lambda_{1}'$, dashed line $\lambda_{2}'$.
Perturbations are not scaled and shown as they are at the simulation
time $\protect\TFinal=4$. Notice, that the left limits of $x$-axes
are taken to be 5 for clarity and are much smaller than the reaction
lengths.}
\label{fig:eigval-pert-sub-sub}
\end{figure}

Now we consider the behavior of the perturbation profiles $z'(x,t)=\{\rho',u',p',\lambda_{1}',\lambda_{2}'\}$
taken at a particular time with $\TFinal=4$. We take $\gamma=1.2$,
$E=15$ and $Q_{1}=\{10,20,50\}$ with $Q_{2}=-0.75Q_{1}$. Figure
\ref{fig:eigval-pert} shows the perturbations for the subsonic-supersonic
case, while Figure \ref{fig:eigval-pert-sub-sub} for the subsonic-subsonic
case. The most prominent distinction between the two figures is that,
while profiles for the subsonic-supersonic case are smooth, profiles
for the subsonic-subsonic case develop discontinuities at the sonic
point caused by discontinuous coefficient matrices $A$ and $B$ in
\eqref{eq:numalg:gov}: ZND profiles themselves are continuous, their
derivatives with respect to $x$ have jump discontinuities at the
sonic point, hence, discontinuous perturbations. In addition, we noticed
that the upwind method (\ref{eq:numal:upwind5-left}\textendash \ref{eq:numalg:upwind5-right})
with Lax\textendash Friedrichs flux led to severe Gibbs oscillations
on both sides of the jump in the profiles of the subsonic-subsonic
case. For this reason, we replaced the upwind method with the WENO
method \cite{JiangPeng2000} for all simulations based on subsonic-subsonic
case. Another important observation from comparing figures \ref{fig:eigval-pert}
and \ref{fig:eigval-pert-sub-sub} is that perturbations are nearly
the same in the region between the sonic point and the leading detonation
shock: for example, the relative difference between density perturbations
is only $1\%$ in this region, and it is likely caused by the smearing
of the discontinuities in the neighborhood of the sonic point.  
 
\subsubsection{Migration of spectra with increasing endothermicity}

Here we investigate how the stability behavior changes as the endothermicity
increases, that is as $Q_{2}$ increases in magnitude while all the
other parameters are kept fixed.

\begin{figure}
\begin{centering}
\includegraphics{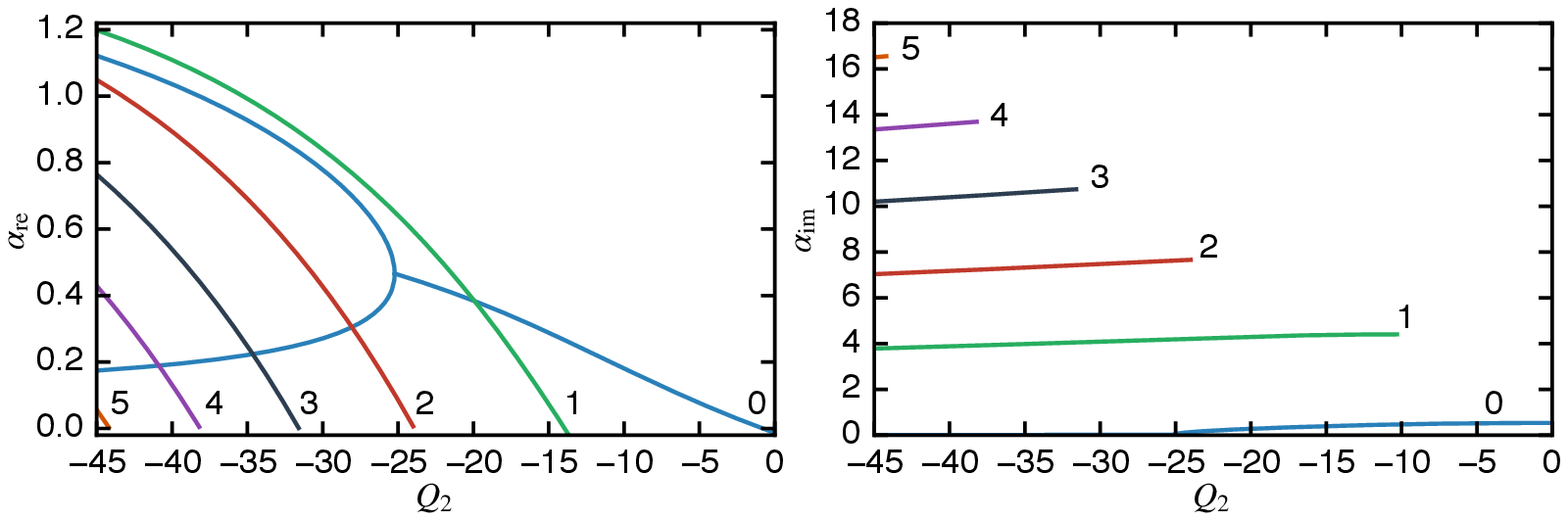}
\par\end{centering}
\caption{Growth rate versus $Q_{2}$ for $-45\leq Q_{2}\leq0$ with other parameters
$\gamma=1.2$, $E=25$, $Q_{1}=50$ for subsonic-supersonic case.\label{fig:eigval-migration-sub-super}}
\end{figure}

First, consider the subsonic-supersonic case. We choose $\Nhrz=40$,
$\gamma=1.2$, $E=25$, $Q_{1}=50$ and let $Q_{2}$ vary in the range
$-45\le Q_{2}\le0$ with step $\Delta Q_{2}=0.0025$ for $Q_{2}\in[-0.7;-0.8]\cup[-13.9;-14]\cup[-19.9;-20]\cup[-25.2;-25.3]$
and step $\Delta Q_{2}=0.1$ otherwise. Figure \ref{fig:eigval-migration-sub-super}
shows how growth rates change as $Q_{2}$ decreases. Initially, at
$Q_{2}=0$ detonation corresponds to the equilibrium CJ case and is
stable for $\gamma$, $E$ and $Q_{1}$ that we use (critical value
of activation energy is $25.26$). As $Q_{2}$ decreases, the growth
rate of the fundamental mode (mode 0) increases almost linearly such
that at $Q_{2}\approx-0.79$ detonation becomes unstable, and at $Q_{2}\approx-25.2225$
mode 0 splits into two purely exponential branches, with growth rates
increasing and decreasing for the top and bottom branches, respectively.
At the same time at $Q_{2}\approx-13.1$ the stable mode 1 appears
in the spectrum, which becomes unstable at $Q_{2}\approx-13.9425$.
As the growth rate of this mode increases faster than that of mode
0, at $Q_{2}\approx-19.92$ it becomes the dominant mode and remains
so all the way down to $Q_{2}=-45$. In total, up to six modes appear
in the spectrum as $Q_{2}\to-45$. Here we point out that taking $|Q_{2}|$
larger than 45 makes computations prohibitively expensive as the effective
reaction lengths become hundreds of thousands of nondimensional units,
because as $|Q_{2}|\to|Q_{1}|$ the overall heat release becomes effectively
zero and detonation degenerates to an inert weak shock.

Next, consider the subsonic-subsonic case. The physical parameters
are taken the same as in the previous case. However, due to the discontinuous
coefficients in the linearized system \eqref{eq:numalg:gov}, now
we must use drastically higher resolutions (here $\Nhrz=1280$) than
for the subsonic-supersonic case. Otherwise, we are unable to extract
low-energy modes. We also emphasize that it is significantly more
difficult to postprocess the time series of perturbation of detonation
velocity for the subsonic-subsonic case. The reason is that the time
series contains substantial amount of numerical noise caused by the
numerical smearing of the jump in the solution in the neighborhood
of the sonic locus. 
\begin{figure}
\begin{centering}
\includegraphics{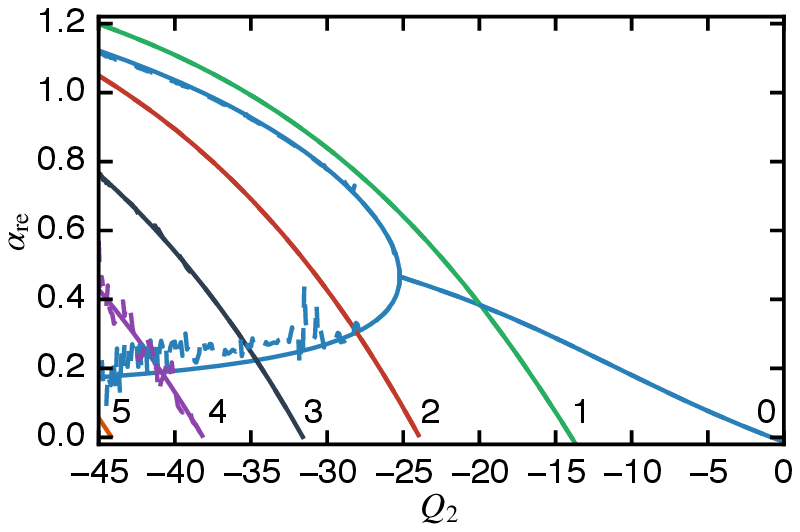}
\par\end{centering}
\caption{Growth rate versus $Q_{2}$ for $-45\leq Q_{2}\leq0$ with $\gamma=1.2$,
$E=25$, $Q_{1}=50$. Solid lines: subsonic-supersonic case, dashed
lines: subsonic-subsonic case. Note that in most cases the solid and
dashed curves overlap and are thus indistinguishable. \label{fig:eigval-re-alpha-vs-q_2-together}}
\end{figure}

Figure \ref{fig:eigval-re-alpha-vs-q_2-together} shows the spectra
for subsonic-supersonic and subsonic-subsonic cases together. It can
be seen that the spectra are close to each other. In fact, they are
essentially indistinguishable for oscillatory part of mode 0, upper
branch of exponential part of mode 0, and for modes 1, 2, and 3. However,
for subsonic-subsonic case, the lower branch of exponential part of
mode 0 shows significant numerical noise as $Q_{2}$ decreases. The
same behavior is seen with mode 4 as well. The reason for this is
that the discontinuity in the solutions for subsonic-subsonic case
generates noise that contaminates time series of $\psi'$. As the
lower branches of modes 0 and 4 contribute to the time series much
less energy than other modes, they are more difficult to distinguish
from the noise, which is an issue that can probably be fixed by using
an improved algorithm for solving hyperbolic problems with discontinuous
coefficients. Nevertheless, the computed values are near their counterparts
for the subsonic-supersonic case. Despite some differences in the
spectra, Figure \ref{fig:eigval-re-alpha-vs-q_2-together} serves
as a demonstration that stability of ZND solutions of the two-step
model with endothermicity does not depend on whether the underlying
ZND solution is of subsonic-supersonic or subsonic-subsonic case.
That is, the detonation stability is apparently determined solely
by the dynamics of the reaction zone between the sonic point and the
leading detonation shock.

As a final note on this section, we give the reader an idea of the
computational requirements for these calculations. For both subsonic-supersonic
and subsonic-subsonic cases, the number of values of $Q_{2}$ we took
is approximately 500. Then, as the resolution necessary for the subsonic-supersonic
case is only $\Nhrz=40$, we were able to compute Figure \ref{fig:eigval-migration-sub-super}
overnight on a workstation with 16 cores. In contrast, the subsonic-subsonic
case requires much higher resolution of $\Nhrz=1280$, so these computations
required a day with 512 parallel workers on a parallel computer, where
a worker simulates and postprocesses results for a given $Q_{2}$
independently of other workers.

\subsubsection{Neutral curves}

In this section, we study how endothermicity affects the neutral stability
boundary in comparison with the one-step model. The comparison is
based on the idea that the detonation dynamics in both cases is determined
by the energy released between the shock and sonic point. Therefore,
the one-step model with heat release $Q$ is compared with the two-step
model with the maximum heat release $Q^{*}$ equal to $Q$. 

\begin{figure}
\begin{minipage}[t]{0.49\columnwidth}%
\includegraphics{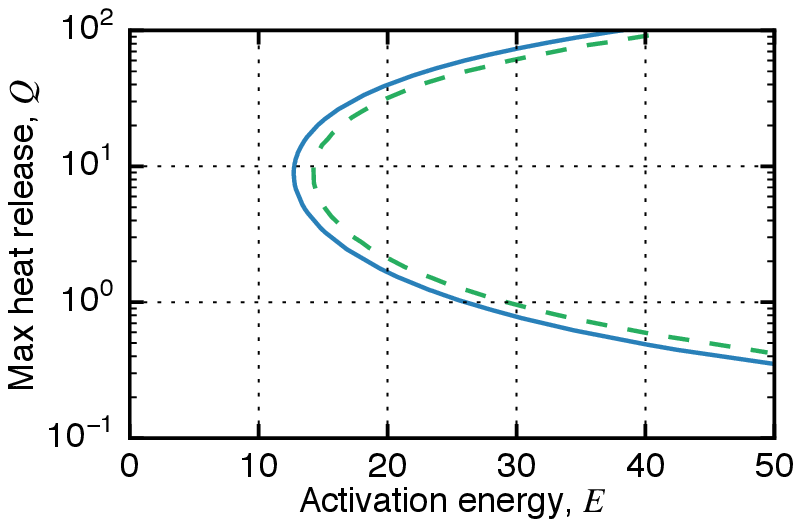}\caption{Neutral stability curves for two-step model with subsonic-supersonic
steady-state solution (solid line) and one-step model (dashed line).\label{fig:eigval-neutral-curve}}
\end{minipage}\hspace*{\fill}%
\begin{minipage}[t]{0.49\columnwidth}%
\includegraphics{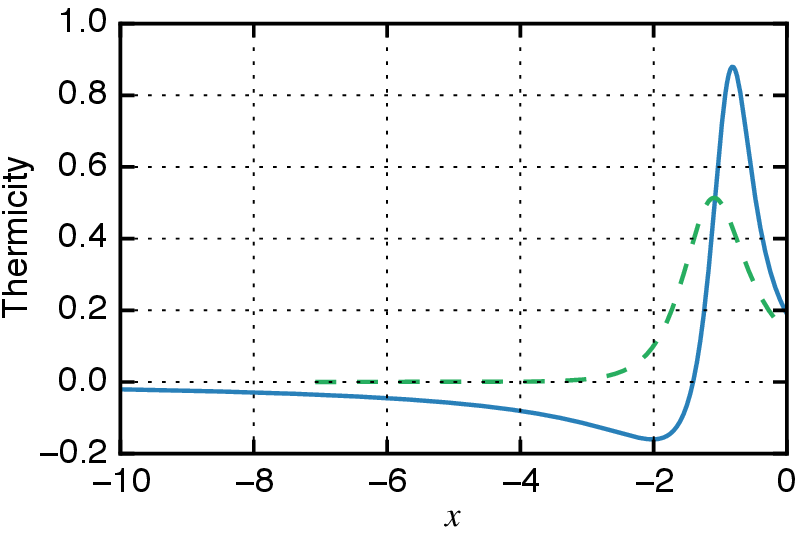}\caption{Thermicity profiles for $Q^{*}=10$, $E=15$ for two-step model with
subsonic-supersonic steady-state solution (solid line) and one-step
model (dashed line).\label{fig:eigval-themicity}}
\end{minipage}
\end{figure}

In Figure \ref{fig:eigval-neutral-curve}, we show the neutral stability
curves in the parametric plane $E$\textendash $Q$. We use $\gamma=1.2$
and therefore the neutral stability curve for the one-step model is
the same as in Figure \ref{fig:LS-comparison}. The neutral curve
for the two-step model is obtained for the ZND solution corresponding
to the subsonic-supersonic case. The range of $Q_{1}$ is $10^{-0.35}$\textendash $10^{2.35}$
with 256 logspaced points while $Q_{2}$ is varied in proportion to
$Q_{1}$: $Q_{2}=-0.75Q_{1}$. For each $Q_{1}$ we find the critical
value of $E$ by the algorithm described earlier. For values of $10^{-0.35}\le Q_{1}\lessapprox0.8$
(approximately 20 points) the algorithm did not converge and we omit
these values from the figure. The curves shown demonstrate that endothermicity
consistently shifts the neutral stability boundary toward smaller
values of $E$ over the whole range of $Q$ that we investigated.
The difference $\Delta E$ between critical activation energies for
two models behaves with varying $Q$ as follows. The minimum value
$\Delta E\approx1.5$ is attained for $5\lessapprox Q\lessapprox15$.
As $Q$ increases, $\Delta E$ increases monotonically and reaches
$\Delta E\approx4.5$ as $Q\to100$. For low values of $Q$, $\Delta E$
increases quickly and in the limit $Q\to0$, $\Delta E\to\infty$
as the detonation wave becomes an inert shock. Additionally, while
for the one-step model, all ZND solutions are stable at $E\leq14.2$
for the full considered range of $Q$, for the two-reaction model
all ZND solutions are stable for $E\leq12.6$.

Thus, the two-step exothermic-endothermic heat release leads to an
increased instability compared to the equivalent one-step heat release.
In order to provide some qualitative explanation for this observation,
we look at how the steady-state thermicity behaves for the same values
of $E$ and $Q^{*}$ for the two models. The thermicity for the two-step
model is given by $\bar{\sigma}_{1}\bar{\omega}_{1}+\bar{\sigma}_{2}\bar{\omega}_{2}$,
where $\bar{\sigma}_{i}=(\gamma-1)Q_{i}/\bar{c}^{2}$, $i=1,2$. Figure
\ref{fig:eigval-themicity} shows the thermicity profiles for $\gamma=1.2$,
$Q^{*}=10$, and $E=15$ which correspond to the unstable steady-state
solutions from Figure \ref{fig:eigval-neutral-curve}. It is clear
that for the two-step model the thermicity has significantly larger
peak value which is in addition located closer to the lead shock than
the value for the one-step model. Thus, a possible reason for the
increased instability is this sharper profile of the thermicity function
that makes for a more effective resonant cavity for the perturbations
between the shock and the sonic point. Recall that the thermicity
is a source term in the equation for the forward characteristics,
$\dot{p}+\rho c\dot{u}=\rho c^{2}\left(\sigma_{1}\omega_{1}+\sigma_{2}\omega_{2}\right)$.
As such it could be looked at as a measure of how the chemical reactions
amplify the forward compression waves that influence the lead shock.
The larger the magnitude of thermicity, the stronger the effect. The
sharpness of the thermicity function reflects in the sensitivity of
this influence on the changes in the local state in the reaction zone,
the stronger sensitivity implying more instability. Obviously, these
are qualitative arguments that require a rational analytical justification
to be seen as a valid explanation, something that is outside the scope
of this work.\selectlanguage{american}%

\section{Conclusions \label{sec:Conclusions}}

In this work, we introduced a new approach to the problem of computing
linear stability properties of detonation waves. The method consists
of two parts. First is a numerical solution of the time-dependent
linearized system of reactive Euler equations by a shock-fitting numerical
algorithm. The result of such numerical calculation is an accurate
time series of the detonation velocity for the linear problem. This
solution is obtained only over relatively short time intervals which
is computationally inexpensive. The second part of the method is an
analysis of the obtained time series through the application of the
method of Dynamic Mode Decomposition. As a result, we are able to
calculate detailed stability properties of gaseous detonations with
either one-step Arrhenius kinetics or two-step kinetics. Adding more
chemical reactions requires essentially the same level of numerical
and analytical effort for the computations. What distinguishes the
proposed method from traditional normal-mode methods for detonations
is that it does not require any special treatment of sonic points
(which requires quite intricate analyses in the normal-mode methods).
The numerical implementation of the method is relatively easy to generalize
to more complex chemistry and equations of state. 

It is worth emphasizing that the methodology we presented in this
work can be applied to any problem of shock wave stability. The key
elements of the analysis (numerical solution of linearized equations
and DMD processing of the computed shock velocity) are clearly independent
of the specific details of the detonation problems that we investigated
in this paper. What is important concerning detonation is that it
possesses strong instabilities and is sensitive to details of numerical
algorithms. As such, detonation represents a very stringent test problem
that is rather demanding of any numerical methods used to compute
its properties. We hope that the present method of stability analysis
can be applied to a wide range of problems where shock wave dynamics
is of importance. 

\vspace{1cm}

\section*{Acknowledgements}

This work was partially supported by the King Abdullah University
of Science and Technology (KAUST) in Thuwal, Saudi Arabia. For computer
time, this work used the resources of the Supercomputing Laboratory
at KAUST. A.\,K. was also partially supported by the Russian Foundation
for Basic Research (grant \#17-53-12018).

\bibliographystyle{abbrv}
\bibliography{kabanov-kasimov-prf-2017.bbl}

\begin{thebibliography}{10}

\bibitem{AbouseifToong82}
G.~Abouseif and T.~Y. Toong.
\newblock Theory of unstable one-dimensional detonations.
\newblock {\em Combust. Flame}, 45:64--94, 1982.

\bibitem{alekseev2016linear}
A.~Alekseev, D.~Bistrian, A.~Bondarev, and I.~Navon.
\newblock On linear and nonlinear aspects of dynamic mode decomposition.
\newblock {\em International Journal for Numerical Methods in Fluids},
  82(6):348--371, 2016.

\bibitem{alla2017nonlinear}
A.~Alla and J.~N. Kutz.
\newblock Nonlinear model order reduction via dynamic mode decomposition.
\newblock {\em SIAM Journal on Scientific Computing}, 39(5):B778--B796, 2017.

\bibitem{arbabi2017study}
H.~Arbabi and I.~Mezi{\'c}.
\newblock {Study of dynamics in unsteady flows using Koopman mode
  decomposition}.
\newblock {\em arXiv preprint arXiv:1704.00813}, 2017.

\bibitem{bagheri2014effects}
S.~Bagheri.
\newblock Effects of weak noise on oscillating flows: linking quality factor,
  floquet modes, and koopman spectrum.
\newblock {\em Physics of Fluids}, 26(9):094104, 2014.

\bibitem{barker2017computing}
B.~Barker, R.~Nguyen, N.~Ventura, and C.~Wahl.
\newblock {Computing Evans functions numerically via boundary-value problems}.
\newblock {\em arXiv preprint arXiv:1710.02500}, 2017.

\bibitem{bates2000d}
J.~W. Bates and D.~C. Montgomery.
\newblock {The D'yakov-Kontorovich instability of shock waves in real gases}.
\newblock {\em Physical Review Letters}, 84(6):1180, 2000.

\bibitem{brown1989vode}
P.~N. Brown, G.~D. Byrne, and A.~C. Hindmarsh.
\newblock {VODE: A variable-coefficient ODE solver}.
\newblock {\em SIAM journal on scientific and statistical computing},
  10(5):1038--1051, 1989.

\bibitem{brunton2016extracting}
B.~W. Brunton, L.~A. Johnson, J.~G. Ojemann, and J.~N. Kutz.
\newblock {Extracting spatial--temporal coherent patterns in large-scale neural
  recordings using dynamic mode decomposition}.
\newblock {\em Journal of neuroscience methods}, 258:1--15, 2016.

\bibitem{brunton2017chaos}
S.~L. Brunton, B.~W. Brunton, J.~L. Proctor, E.~Kaiser, and J.~N. Kutz.
\newblock Chaos as an intermittently forced linear system.
\newblock {\em Nature Communications}, 8, 2017.

\bibitem{chen2012variants}
K.~K. Chen, J.~H. Tu, and C.~W. Rowley.
\newblock {Variants of dynamic mode decomposition: boundary condition, Koopman,
  and Fourier analyses}.
\newblock {\em Journal of Nonlinear Science}, 22(6):887--915, 2012.

\bibitem{dawson2016characterizing}
S.~T.~M. Dawson, M.~S. Hemati, M.~O. Williams, and C.~W. Rowley.
\newblock Characterizing and correcting for the effect of sensor noise in the
  dynamic mode decomposition.
\newblock {\em Experiments in Fluids}, 57(3):1--19, 2016.

\bibitem{Doering1943}
W.~D{\"o}ring.
\newblock Uber den detonationvorgang in gasen.
\newblock {\em Annalen der Physik}, 43(6/7):421--428, 1943.

\bibitem{drazin2004hydrodynamic}
P.~G. Drazin and W.~H. Reid.
\newblock {\em Hydrodynamic stability}.
\newblock Cambridge university press, 2004.

\bibitem{duke2012error}
D.~Duke, J.~Soria, and D.~Honnery.
\newblock An error analysis of the dynamic mode decomposition.
\newblock {\em Experiments in Fluids}, 52(2):529--542, 2012.

\bibitem{dyakov1954}
S.~P. D'yakov.
\newblock The stability of shock waves.
\newblock {\em Journal of Experimental and Theoretical Physics},
  27(3):288--296, 1954.

\bibitem{el2016dispersive}
G.~A. El and M.~A. Hoefer.
\newblock Dispersive shock waves and modulation theory.
\newblock {\em Physica D: Nonlinear Phenomena}, 333:11--65, 2016.

\bibitem{erpenbeck1962stability}
J.~J. Erpenbeck.
\newblock Stability of step shocks.
\newblock {\em The Physics of Fluids}, 5(10):1181--1187, 1962.

\bibitem{Erpenbeck64}
J.~J. Erpenbeck.
\newblock Stability of idealized one-reaction detonations.
\newblock {\em Phys. Fluids}, 7:684--696, 1964.

\bibitem{fernandez2009stability}
R.~Fern{\'a}ndez and C.~Thompson.
\newblock Stability of a spherical accretion shock with nuclear dissociation.
\newblock {\em The Astrophysical Journal}, 697(2):1827, 2009.

\bibitem{FickettDavis2011}
W.~Fickett and W.~C. Davis.
\newblock {\em Detonation: theory and experiment}.
\newblock Dover Publications, 2011.

\bibitem{foglizzo2007instability}
T.~Foglizzo, P.~Galletti, L.~Scheck, and H.-T. Janka.
\newblock Instability of a stalled accretion shock: evidence for the
  advective-acoustic cycle.
\newblock {\em The Astrophysical Journal}, 654(2):1006, 2007.

\bibitem{fowles1973stability}
G.~R. Fowles and G.~W. Swan.
\newblock Stability of plane shock waves.
\newblock {\em Physical Review Letters}, 30(21):1023, 1973.

\bibitem{golyandina2013singular}
N.~Golyandina and A.~Zhigljavsky.
\newblock {\em Singular Spectrum Analysis for time series}.
\newblock {Springer Science \& Business Media}, 2013.

\bibitem{hairer1993solving}
E.~Hairer, S.~P. N{\o}rsett, and G.~Wanner.
\newblock {\em {Solving Ordinary Differential Equations I: Nonstiff problems}}.
\newblock Springer, 1993.

\bibitem{HenrickAslamPowers2006}
A.~K. Henrick, T.~D. Aslam, and J.~M. Powers.
\newblock Simulations of pulsating one-dimensional detonations with true fifth
  order accuracy.
\newblock {\em J. Comput. Phys.}, 213(1):311--329, 2006.

\bibitem{humpherys2010efficient}
J.~Humpherys and K.~Zumbrun.
\newblock Efficient numerical stability analysis of detonation waves in {ZND}.
\newblock {\em arXiv preprint arXiv:1011.0897}, 2010.

\bibitem{JiangPeng2000}
G.~Jiang and D.~Peng.
\newblock {Weighted ENO Schemes for Hamilton-Jacobi Equations}.
\newblock {\em SIAM J. Sci. Comput.}, 21(6):2126--2143, 2000.

\bibitem{jovanovic2014sparsity}
M.~R. Jovanovi{\'c}, P.~J. Schmid, and J.~W. Nichols.
\newblock Sparsity-promoting dynamic mode decomposition.
\newblock {\em Physics of Fluids}, 26(2):024103, 2014.

\bibitem{Kasimov-phd}
A.~R. Kasimov.
\newblock {\em Theory of instability and nonlinear evolution of self-sustained
  detonation waves}.
\newblock PhD thesis, University of Illinois at Urbana-Champaign, Urbana, IL,
  USA, 2004.

\bibitem{KasimovStewart02}
A.~R. Kasimov and D.~S. Stewart.
\newblock Spinning instability of gaseous detonations.
\newblock {\em J. Fluid Mech.}, 466:179--203, 2002.

\bibitem{kasimov2004dynamics}
A.~R. Kasimov and D.~S. Stewart.
\newblock On the dynamics of self-sustained one-dimensional detonations: A
  numerical study in the shock-attached frame.
\newblock {\em Physics of Fluids}, 16:3566, 2004.

\bibitem{KasimovStewart05}
A.~R. Kasimov and D.~S. Stewart.
\newblock Asymptotic theory of evolution and failure of self-sustained
  detonations.
\newblock {\em J. Fluid Mech.}, 525:161--192, 2005.

\bibitem{kim2014nature}
W.-T. Kim, Y.~Kim, and J.-G. Kim.
\newblock Nature of the wiggle instability of galactic spiral shocks.
\newblock {\em The Astrophysical Journal}, 789(1):68, 2014.

\bibitem{kontorovich1958concerning}
V.~M. Kontorovich.
\newblock Concerning the stability of shock waves.
\newblock {\em Soviet Phys. JETP}, 6, 1958.

\bibitem{koopman1931hamiltonian}
B.~O. Koopman.
\newblock Hamiltonian systems and transformation in hilbert space.
\newblock {\em Proceedings of the National Academy of Sciences},
  17(5):315--318, 1931.

\bibitem{kou2017improved}
J.~Kou and W.~Zhang.
\newblock An improved criterion to select dominant modes from dynamic mode
  decomposition.
\newblock {\em European Journal of Mechanics-B/Fluids}, 62:109--129, 2017.

\bibitem{kutz2016dynamic}
J.~N. Kutz, S.~L. Brunton, B.~W. Brunton, and J.~L. Proctor.
\newblock {\em {Dynamic Mode Decomposition: Data-Driven Modeling of Complex
  Systems}}.
\newblock SIAM, 2016.

\bibitem{kutz2016koopman}
J.~N. Kutz, J.~L. Proctor, and S.~L. Brunton.
\newblock Koopman theory for partial differential equations.
\newblock {\em arXiv preprint arXiv:1607.07076}, 2016.

\bibitem{kuznetsov1989stability}
N.~Kuznetsov.
\newblock Stability of shock waves.
\newblock {\em Physics-Uspekhi}, 32(11):993--1012, 1989.

\bibitem{laney1998}
C.~B. Laney.
\newblock {\em Computational gasdynamics}.
\newblock Cambridge University Press, 1998.

\bibitem{LeeStewart90}
H.~I. Lee and D.~S. Stewart.
\newblock Calculation of linear detonation instability: One-dimensional
  instability of plane detonation.
\newblock {\em J. Fluid Mech.}, 212:103--132, 1990.

\bibitem{massa2012dynamic}
L.~Massa, R.~Kumar, and P.~Ravindran.
\newblock Dynamic mode decomposition analysis of detonation waves.
\newblock {\em Physics of Fluids}, 24(6):066101, 2012.

\bibitem{mezic2013analysis}
I.~Mezi{\'c}.
\newblock {Analysis of fluid flows via spectral properties of the Koopman
  operator}.
\newblock {\em Annual Review of Fluid Mechanics}, 45:357--378, 2013.

\bibitem{peitz2017koopman}
S.~Peitz and S.~Klus.
\newblock {Koopman operator-based model reduction for switched-system control
  of PDEs}.
\newblock {\em arXiv preprint arXiv:1710.06759}, 2017.

\bibitem{rezzolla1996stability}
L.~Rezzolla.
\newblock Stability of cosmological detonation fronts.
\newblock {\em Physical Review D}, 54(2):1345, 1996.

\bibitem{richecoeur2012dmd}
F.~Richecoeur, L.~Hakim, A.~Renaud, and L.~Zimmer.
\newblock {DMD algorithms for experimental data processing in combustion},
  2012.

\bibitem{rowley2017model}
C.~W. Rowley and S.~T. Dawson.
\newblock Model reduction for flow analysis and control.
\newblock {\em Annual Review of Fluid Mechanics}, 49:387--417, 2017.

\bibitem{rowley2009spectral}
C.~W. Rowley, I.~Mezi{\'c}, S.~Bagheri, P.~Schlatter, and D.~S. Henningson.
\newblock Spectral analysis of nonlinear flows.
\newblock {\em Journal of Fluid Mechanics}, 641:115--127, 2009.

\bibitem{schmid2010dynamic}
P.~J. Schmid.
\newblock Dynamic mode decomposition of numerical and experimental data.
\newblock {\em Journal of Fluid Mechanics}, 656:5--28, 2010.

\bibitem{schmid2011application}
P.~J. Schmid.
\newblock Application of the dynamic mode decomposition to experimental data.
\newblock {\em Experiments in Fluids}, 50(4):1123--1130, 2011.

\bibitem{schmid2012stability}
P.~J. Schmid and D.~S. Henningson.
\newblock {\em Stability and transition in shear flows}, volume 142.
\newblock {Springer Science \& Business Media}, 2012.

\bibitem{schmid2011applications}
P.~J. Schmid, L.~Li, M.~P. Juniper, and O.~Pust.
\newblock Applications of the dynamic mode decomposition.
\newblock {\em Theoretical and Computational Fluid Dynamics}, 25(1-4):249--259,
  2011.

\bibitem{schmid2008decomposition}
P.~J. Schmid and J.~L. Sesterhenn.
\newblock Dynamic mode decomposition of numerical and experimental data.
\newblock In {\em 61st Annual Meeting of the APS Division of Fluid Dynamics.}
  American Physical Society, November 2008.

\bibitem{sharpe1997linear}
G.~Sharpe.
\newblock Linear stability of idealized detonations.
\newblock {\em Proceedings of the Royal Society of London. Series A:
  Mathematical, Physical and Engineering Sciences}, 453(1967):2603, 1997.

\bibitem{sharpe2000one}
G.~Sharpe and S.~Falle.
\newblock One-dimensional nonlinear stability of pathological detonations.
\newblock {\em Journal of Fluid Mechanics}, 414:339--366, 2000.

\bibitem{sharpe1999linearpathological}
G.~J. Sharpe.
\newblock Linear stability of pathological detonations.
\newblock {\em Journal of Fluid Mechanics}, 401:311--338, 1999.

\bibitem{short1998cellular}
M.~Short and D.~S. Stewart.
\newblock {Cellular detonation stability. Part 1. A normal-mode linear
  analysis}.
\newblock {\em Journal of Fluid Mechanics}, 368:229--262, 1998.

\bibitem{StewartKasimovSIAP05}
D.~S. Stewart and A.~R. Kasimov.
\newblock Theory of detonation with an embedded sonic locus.
\newblock {\em SIAM J. Appl. Maths.}, 66(2):384--407, 2005.

\bibitem{swan1975shock}
G.~W. Swan and G.~R. Fowles.
\newblock Shock wave stability.
\newblock {\em The Physics of Fluids}, 18(1):28--35, 1975.

\bibitem{syrovatskii1959stability}
S.~I. Syrovatskii.
\newblock The stability of shock waves in magnetohydrodynamics.
\newblock {\em Soviet Phys. JETP}, 35:1024--1027, 1959.

\bibitem{taira2017modal}
K.~Taira, S.~L. Brunton, S.~Dawson, C.~W. Rowley, T.~Colonius, B.~J. McKeon,
  O.~T. Schmidt, S.~Gordeyev, V.~Theofilis, and L.~S. Ukeiley.
\newblock {Modal analysis of fluid flows: An overview}.
\newblock {\em arXiv preprint arXiv:1702.01453}, 2017.

\bibitem{Taylor-Kasimov-Stewart-CTM09}
B.~D. Taylor, A.~R. Kasimov, and D.~S. Stewart.
\newblock Mode selection in unstable two-dimensional detonations.
\newblock {\em Combust. Theory Model}, 13(6):973--992, 2009.

\bibitem{tu2014dynamic}
J.~H. Tu, C.~W. Rowley, D.~M. Luchtenburg, S.~L. Brunton, and J.~N. Kutz.
\newblock {On dynamic mode decomposition: theory and applications}.
\newblock {\em Journal of Computational Dynamics}, 1(2):391--421, 2014.

\bibitem{tumin2007initial}
A.~Tumin.
\newblock Initial-value problem for small disturbances in an idealized
  one-dimensional detonation.
\newblock {\em Physics of Fluids}, 19:106105, 2007.

\bibitem{vonNeumann1942}
J.~von Neumann.
\newblock Theory of detonation waves. {O}ffice of {S}cientific {R}esearch and
  {D}evelopment, {R}eport 549.
\newblock Technical report, National Defense Research Committee Div. B, 1942.

\bibitem{williams2015data}
M.~O. Williams, I.~G. Kevrekidis, and C.~W. Rowley.
\newblock {A data-driven approximation of the Koopman operator: Extending
  dynamic mode decomposition}.
\newblock {\em Journal of Nonlinear Science}, 25(6):1307--1346, 2015.

\bibitem{wynn2013optimal}
A.~Wynn, D.~Pearson, B.~Ganapathisubramani, and P.~J. Goulart.
\newblock Optimal mode decomposition for unsteady flows.
\newblock {\em Journal of Fluid Mechanics}, 733:473--503, 2013.

\bibitem{Zeldovich1940}
Y.~B. Zel'dovich.
\newblock On the theory of propagation of detonation in gaseous systems.
\newblock {\em J. Exp. Theor. Phys.}, 10(5):542--569, 1940.

\bibitem{zhang2017evaluating}
H.~Zhang, S.~Dawson, C.~W. Rowley, E.~A. Deem, and L.~N. Cattafesta.
\newblock Evaluating the accuracy of the dynamic mode decomposition.
\newblock {\em arXiv preprint arXiv:1710.00745}, 2017.

\bibitem{zumbrun2011stability}
K.~Zumbrun.
\newblock Stability of detonation profiles in the {ZND} limit.
\newblock {\em Archive for rational mechanics and analysis}, 200(1):141--182,
  2011.

\bibitem{zumbrun2017recent}
K.~Zumbrun.
\newblock Recent results on stability of planar detonations.
\newblock In {\em Shocks, Singularities and Oscillations in Nonlinear Optics
  and Fluid Mechanics}, pages 273--308. Springer, 2017.

\end{thebibliography}

\end{document}